\documentclass[aps,prl,twocolumn,amsmath,amssymb,superscriptaddress]{revtex4-1}
\usepackage[english]{babel}
\usepackage{amsmath}
\usepackage{amsfonts}
\usepackage{graphicx}
\usepackage{color}

\newcommand{\beq}{\begin{equation}} 
\newcommand{\eeq}{\end{equation}}

\begin{document}
\title {Dynamic current susceptibility as a probe of Majorana bound states in nanowire-based Josephson junctions}
\author{Mircea Trif}
\affiliation{Institute for Interdisciplinary Information Sciences, Tsinghua University, Beijing, China}

\author{Olesia Dmytruk}
\affiliation{Laboratoire de Physique des Solides, CNRS, Univ. Paris-Sud,
Universit\'e Paris Saclay, 91405 Orsay cedex, France}
\affiliation{Department of Physics, University of Basel, Klingelbergstrasse 82, CH-4056 Basel, Switzerland}

\author{Helene Bouchiat}
\affiliation{Laboratoire de Physique des Solides, CNRS, Univ. Paris-Sud,
Universit\'e Paris Saclay, 91405 Orsay cedex, France}

\author{Ram\'on Aguado}
\affiliation{Instituto de Ciencia de Materiales de Madrid (ICMM),
Consejo Superior de Investigaciones Cient\'ificas (CSIC),
Sor Juana In\'es de la Cruz 3, 28049 Madrid, Spain}

\author{Pascal Simon}
\affiliation{Laboratoire de Physique des Solides, CNRS, Univ. Paris-Sud,
Universit\'e Paris Saclay, 91405 Orsay cedex, France}
\date{\today}

\begin{abstract}
We theoretically study a Josephson junction based on a semiconducting nanowire subject to a time-dependent flux bias. We establish a general density matrix approach for the dynamical response of the Majorana junction and calculate the resulting flux-dependent susceptibility using both microscopic and effective low-energy descriptions for the nanowire. We find that the diagonal component of the  susceptibility, associated with the dynamics of the Majorana states populations, dominates over the standard Kubo contribution for a wide range of experimentally relevant parameters. The diagonal term, thus far unexplored in the context of Majorana physics, allows to probe accurately the presence of Majorana bound states in the junction. 
\end{abstract}


\maketitle

\emph{Introduction---} Majorana bound states (MBS) are zero-energy Bogoliubov-de Gennes (BdG) quasiparticles in so-called topological superconductors. They exhibit non-Abelian
 exchange statistics that makes them attractive as building blocks for a fault-tolerant topological quantum computer~\cite{kitaev2001unpaired,dassarma-majarona-review,aasen2016-qc}. This technological potential, together with their intrinsic fundamental interest, has motivated a great deal of excitement towards detecting and manipulating MBS in various condensed matter platforms~\cite{alicea2012new,franz2015review}. 

Arguably, the platform that has attracted the most excitement is the one based on one-dimensional (1D) semiconducting wires (SW). 
Following theoretical proposals~\cite{lutchyn2010majorana,oreg2010helical}, several experiments~\cite{mourik2012signatures,xu2012,Das2012,churchill2013,albrecht2016exponential,Deng2016,Kouwenhoven2016,albrecht2016transport} have reported characteristic transport signatures in the form of a zero-bias conductance peak compatible with the presence of zero-energy MBS. 
Despite this evidence, however, the nagging question of whether zero-bias peaks are due to MBS is still under debate~\cite{dassarma2017}.
Therefore, it would be very useful to study alternative signatures of MBS beyond zero-bias peaks.

\begin{figure}[t] 
\centering
\includegraphics[width=0.6\linewidth]{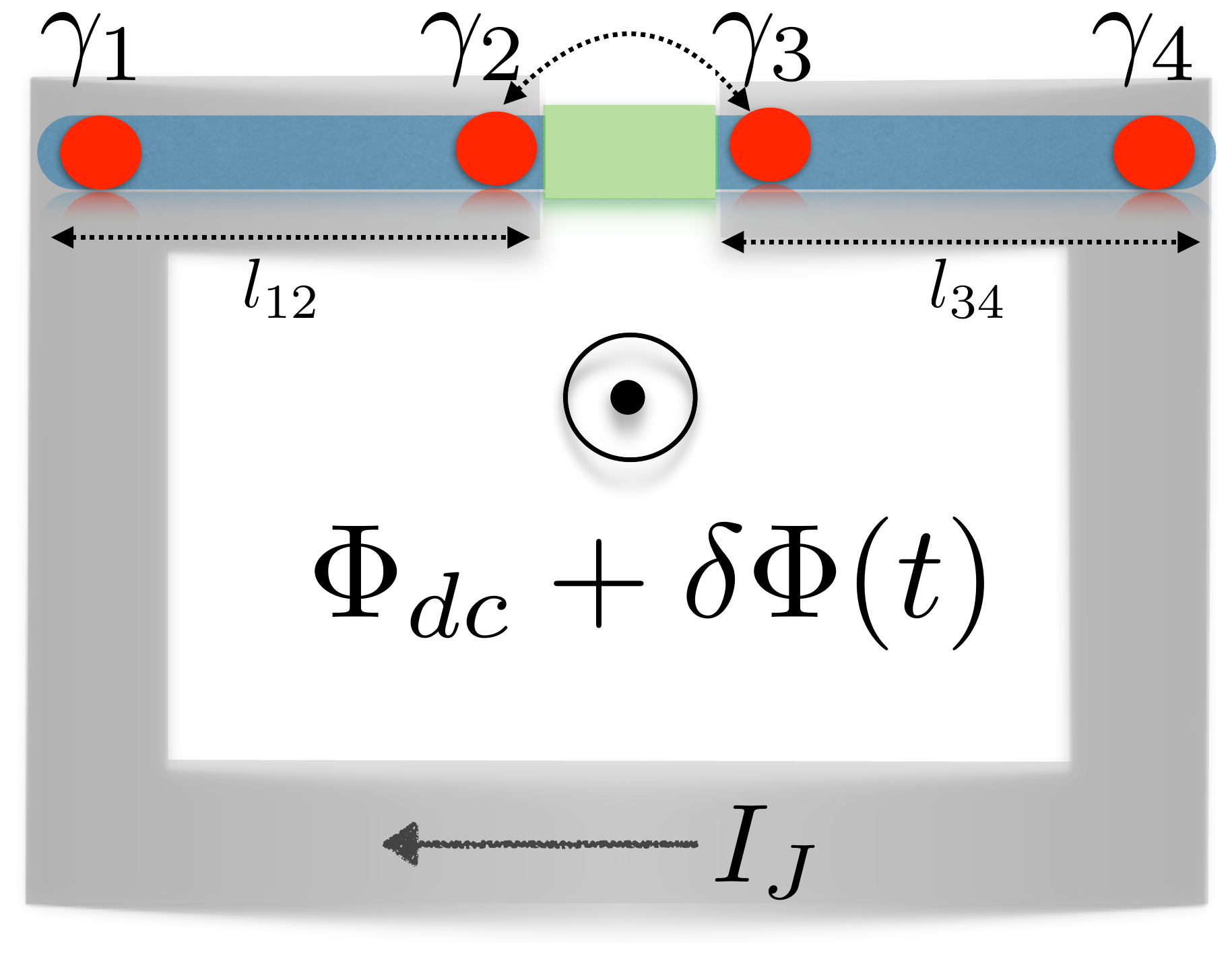}
\caption{Scheme of the setup: A nanowire (in blue) proximitized with bulk s-wave superconductor (in grey) and subject to both a dc 
flux, $\Phi_{dc}$, and an ac flux $\Phi_{ac}=\delta{\Phi}(t)$. The superconducting phase across the junction and the external flux threading the ring are related by the condition $\phi=2e\Phi/\hbar$, with $\Phi=\Phi_{dc}+\delta\Phi$.
 The bulk s-wave superconductor (in grey) is interrupted under the weak link (in green). Red circles indicate four MBS $\gamma_1 - \gamma_4$, while $l_{12}$ and $l_{34}$ are the sizes of the left and right topological regions, respectively. The two inner MBS $\gamma_2$ and $\gamma_3$ overlap through the weak link 
while the outer MBS $\gamma_1$ and $\gamma_4$ can in principle overlap through the superconducting ring. 
} 
\label{fig:scheme} 
\end{figure}

One option is to study the $4\pi$-periodic Josephson effect in junctions based on topological SWs~\cite{kitaev2001unpaired,kwon2004fractional,fu2009josephson}. 
This seemingly smoking-gun signature of MBS in the junction is, however, not free from drawbacks: either parity breaking processes, such as 
quasiparticle poisoning, or the intrinsic avoided crossing of MBS in a realistic finite-size topological SW will unavoidably restore the $2\pi$ periodicity of the ground state~\cite{Nazarov-pikulin2011,san2012ac}.  In principle, one can restore the $4\pi$-periodicity by voltage-biasing the junction and studying the ac Josephson effect~\cite{Nazarov-pikulin2011,san2012ac,Badiane2011}. However, various non-equilibrium and non-adiabatic effects, such as Landau-Zenner tunneling (LZT), make the interpretation in terms of MBS challenging.

We here propose to overcome these difficulties by focusing on a fully linear response quantity: the finite frequency current susceptibility of the junction $\chi(\phi,\omega)=i\omega Y(\phi,\omega)$, where $Y(\phi,\omega)$ is the admittance.  $\chi(\phi,\omega)$ can be obtained by adding a small ac component to the superconducting phase $\phi\rightarrow\phi+\delta\phi(t)$. This can be accomplished by inserting the SW in a superconducting ring geometry (Fig.~\ref{fig:scheme}) that is measured by a microwave resonator~\cite{chiodi2011,dassonneville2013dissipation}. 
Such a scheme avoids unwanted complications, such as LZT, owing to the intrinsic near equilibrium character of the ac phase biasing \cite{NoteA}. 

\begin{figure}[t] 
\centering
\includegraphics[width=0.9\linewidth]{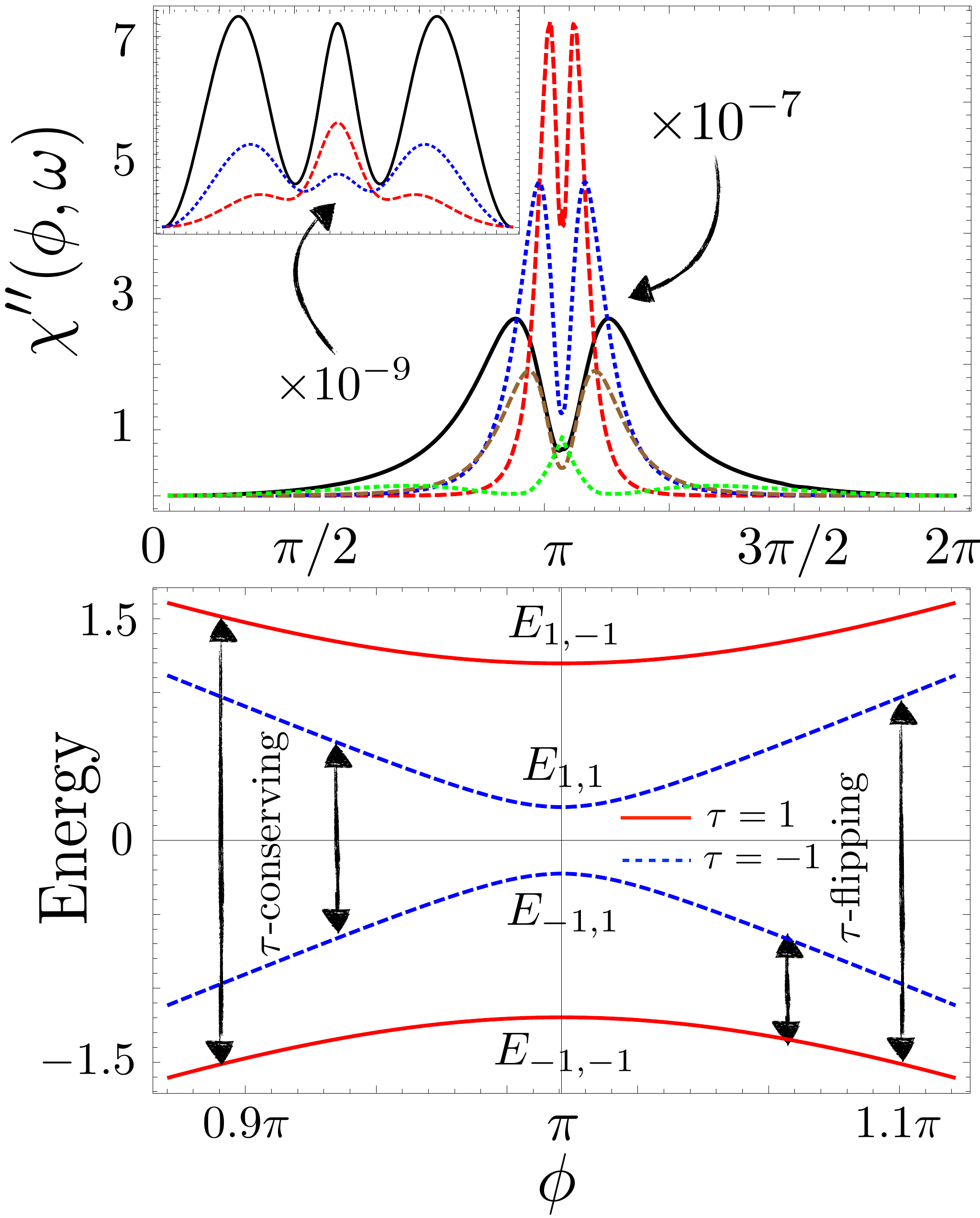}\
\caption{Top: The imaginary part of the susceptibility $\chi^{''}(\phi,\omega)$ as a function of the phase $\phi$ in the topological regime for various values of the Zeeman field and assuming all possible transitions. The black, red, blue, brown, and  green  lines correspond to $V_Z=1.2, 1.3, 1.4, 1.5$, and  $1.6$ ($\times\,\Delta_s^w$), respectively  [Inset: The  black, red, blue lines correspond to $V_Z=1.7, 2$, and  $2.3$, ($\times\,\Delta_s^w$) respectively].  The topological transition takes place at $V_z=\Delta_s^w$. We expressed all energies in terms of the hopping $t$, with $\Delta_s^w=0.05$, $\alpha=0.08$, $\omega=1.6\times10^{-4}$, $T=0.8\,\omega$, $\gamma=10^{-8}$. Bottom:  Many-body spectrum and the allowed transitions for $V_z=1.2\Delta_s^w$ (all other parameters as above). The red (blue) levels correspond to the $\tau=1$ ($\tau=-1$) parity state. Left (right) vertical arrows depict the parity conserving (flipping) transitions. } 
\label{Zeeman} 
\end{figure}

\begin{figure}[t] 
\centering
\includegraphics[width=0.9\linewidth]{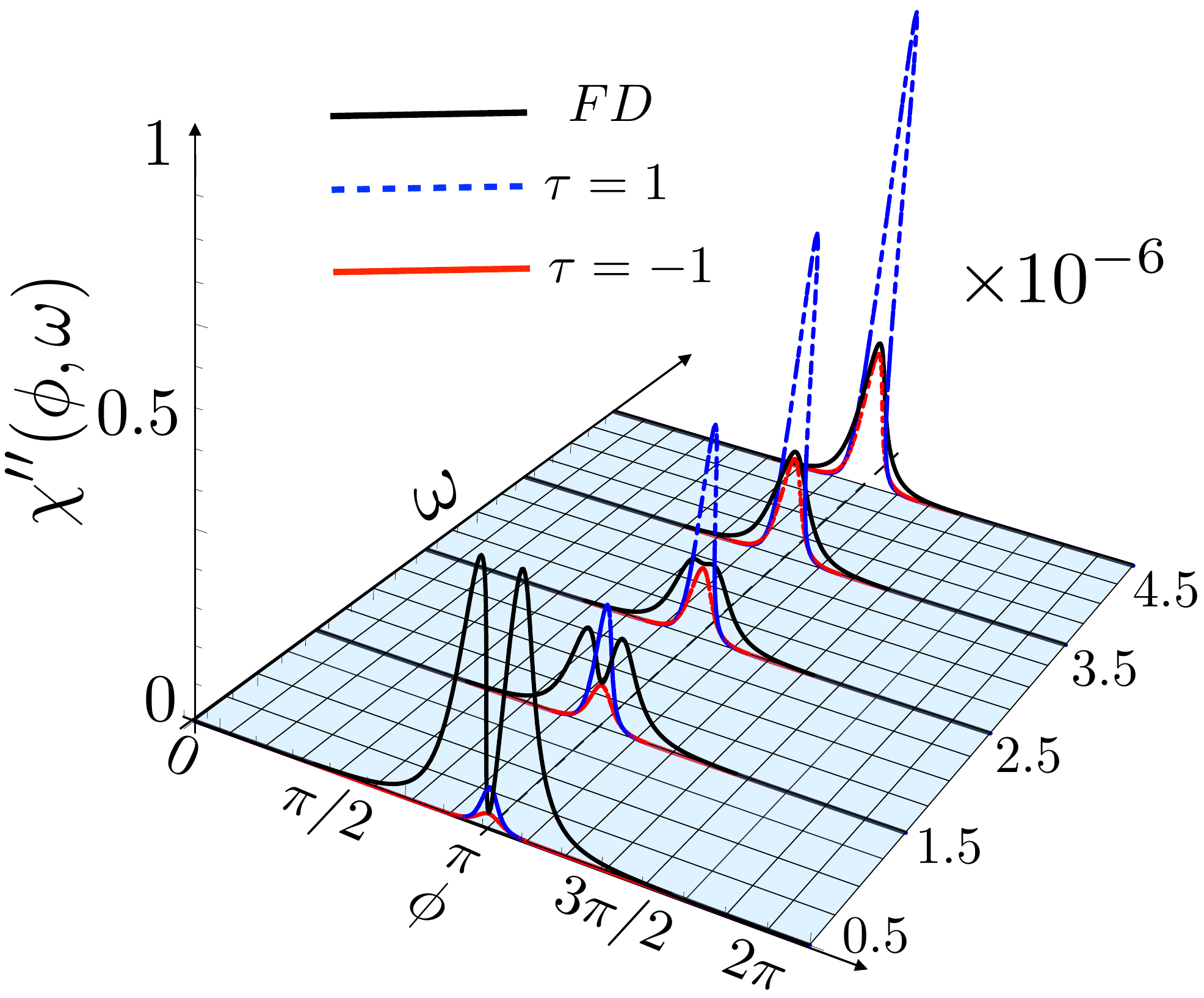}
\caption{The imaginary part of the total susceptibility $\chi{''}(\phi,\omega)$ from the effective model 
 as a function of $\phi$
with and without the parity constraints for several values of the resonator frequency $\omega$ (relative to the frequency in Fig.~2).  The black, red, and blue curves correspond to unconstrained (Fermi-Dirac) distributions, constrained with parity $\tau=1$, and $\tau=-1$, respectively. The parameters of the effective model are extracted from the full Hamiltonian at $V_z=1.5$, while the rest of the parameters are the same as in Fig.~\ref{Zeeman}. A crossover from  $\chi''_{D}$ to $\chi''_{ND}$  dominated dissipation is clearly visible in the range of frequencies depicted in the plot for the unconstrained parity case, while the dissipation is dominated by $\chi''_{\tau,ND}$ in the constrained case.} 
\label{Parity} 
\end{figure}

In particular, we focus here on the first dynamic correction beyond the purely static contribution to the Josephson inductance $L_J^{-1}(\phi)=\chi_J(\phi)=\frac{2e}{\hbar}\partial_\phi J_s(\phi)$, where $J_s(\phi)$ is the Josephson current. This low-frequency contribution $\chi_D(\phi,\omega)$, see Eq.~(\ref{eq:suscept}), which physically originates from the dynamics of populations of excited Andreev levels in the junction  remains, to the best of our knowledge, hitherto unexplored in the context of Majorana wires. This is in contrast to the high-frequency susceptibility, $\chi_{ND}(\phi,\omega)$ in Eq.~(\ref{eq:suscept}), whose imaginary part directly describes microwave-induced transitions between Andreev levels \cite{tewari2012probing,vayrynen2015microwave,peng2016signatures,dmytruk2016josephson}.  Our main results are summarized in Figs.~\ref{Zeeman} and \ref{Parity}, where we demonstrate that, $\chi_D(\phi,\omega)$, as a low-frequency accessible quantity, contains unique signatures due to MBS in the junction. Furthermore, we show that such quantity is sensitive to the parity distribution function (see Fig.~\ref{Parity}) and therefore allows to estimate the parity lifetime at the probe frequency.

\emph{System and Hamiltonian--} The SW  is composed of three parts, a left (L) and right (R) superconducting part (in blue in Fig.~\ref{fig:scheme}) and a normal (No) central part (in green in Fig.~\ref{fig:scheme}). The  Hamiltonian describing the SW reads~\cite{rainis2013towards}, $H_{\rm w} = H_{L} + H_{R} + H_{No}$, [see the Supplemental Material (SM)~\cite{sm}] where $H_{s=L,R,No}$  is given by
\begin{align}
H_{s}&=\sum_{j\in s; \sigma,\sigma'}\Big[-c^\dag_{j+1,\sigma}(t\delta_{\sigma\sigma'}+i\alpha\sigma^y_{\sigma\sigma'})c_{j,\sigma'}+{\rm H. c. }\nonumber\\
&- c^\dag_{j,\sigma}(\mu\delta_{\sigma\sigma'}+V_z\sigma^x_{\sigma\sigma'})c_{j,\sigma'}+\Delta^s_w c^\dag_{j,\uparrow}c^\dag_{j,\downarrow}\Big]\,.
\label{eq:wire}
\end{align}
Here,
 $t$ is the hopping amplitude, $\mu$ is the chemical potential,  $\alpha$ is the spin-flip hopping amplitude, $\Delta_w^s$ is the pairing potential proximity induced from the superconductor, $V_z$ is the Zeeman energy ($V_z=g\mu_BB/2$) and $\sigma^{x,y}$  are Pauli matrices. Also, $c^\dagger_{j\sigma}$ ($c_{j\sigma}$) are the fermionic creation (annihilation) operators at site $j$ and for spin $\sigma$. We mention that the pairing is induced into the wire by the nearby $s$-wave superconductors via the proximity effect.
The many-body Hamiltonian can be written as $H_{w}=\dfrac{1}{2}\vec{c}^\dagger H_{BdG}\vec{c}$, with $H_{BdG}$ being the BdG Hamiltonian describing the single-particle excitations, and written in the basis $\vec{c}^\dagger\equiv(c^\dagger_{1\uparrow}, c^\dagger_{1\downarrow},\dots c^\dagger_{N_w\uparrow}, c^\dagger_{N_w\downarrow},c_{1\uparrow}, c_{1\downarrow},\dots, c_{N_w\uparrow},c_{N_w\downarrow})$, with $N_w$ the total number of sites in the wire.

To complete our setup, we assume the ring is inductively coupled to a microwave superconductor resonator and also threaded by a dc magnetic flux $\Phi_{dc}$, so that the total flux is $\Phi(t)=\Phi_{dc}+\delta\Phi(t)$, which effectively acts as to induce a phase difference between the superconducting pairing across the normal link, {\it i.e.} $\Delta_w^{R}=\Delta_w e^{i\phi}$, $\Delta_w^{L}=\Delta_w$ and 
$\Delta_w^{No}=0$.  We mention that for $V_z^s > \sqrt{(\mu_s+2t_s)^2 + (\Delta_w^s)^2}$, there are four MBS present in the ring: $\gamma_1$, $\gamma_4$ localized at the ends of the spin-orbit coupled nanowire, and $\gamma_2$, $\gamma_3$ localized on both sides of the weak link (see Fig.~\ref{fig:scheme}).  Otherwise, no MBS emerge and the system is in the topologically trivial phase.

\emph{Density matrix and evolution --} 
In the presence of the ac flux, such that $\delta\Phi\ll\Phi_{dc}$, the Hamiltonian can be written as $H_{BdG}(t)=H_{BdG}+V_{BdG}(t)$, with $H_{BdG}$ the Hamiltonian in the absence of the ac flux~\cite{vayrynen2015microwave}, and 
$ V_{BdG}(t)=-\delta\Phi(t) \hat{I}_s$.
Here, $\hat{I}_s\equiv -\partial H_{BdG}/\partial\Phi$ is the current operator in the absence of the perturbation. 
The time-dependent system is described by the following density matrix evolution~\cite{trivedi1988mesoscopic,ferrier2013phase,sticlet2014dynamical,dmytruk2016josephson,murani2016andreev}:
\begin{equation}
\frac{\partial \rho_{BdG}(t)}{\partial t}+\dfrac{i}{\hbar}[H_{BdG}(t),\rho(t)]=-\hat{\Gamma}[\rho_{BdG}(t)-\rho_{BdG,qe}(t)]\,,
\end{equation}
where $\rho_{BdG}(t)$ denotes the reduced density matrix of the system (after tracing over the environment), $\hat{\Gamma}$ is the reduced relaxation tensor (that accounts for both the diagonal and off-diagonal relaxations)  and $\rho_{BdG,qe}(t)$ is the (time-dependent) quasi-equilibrium density matrix. If the parity of the system is not constrained, this is just the Fermi-Dirac (FD) distribution.  Otherwise, the distribution needs to be evaluated subject to constraints, an issue we will describe  further. The time-dependent average current can be found from $\langle\hat{I}_s(t)\rangle={\rm Tr}[\hat{I}_s(t)\rho_{BdG}(t)]$, with $\hat{I}_s(t)=-\partial H_{BdG}(t)/\partial\Phi(t)$, the current operator in the presence of the driving field. One defines the susceptibility of the system as $\chi(\phi)=\delta\langle\hat{I}_s(t)\rangle/\delta\Phi(t)$, with $\delta\langle\hat{I}_s(t)\rangle\equiv\langle\hat{I}_s(t)\rangle-{\rm Tr}[\hat{I}_s\rho^0_{BdG,qe}]$ being the deviation of the curent in the presence of the drive from the equilibrium current. As shown previously~\cite{trivedi1988mesoscopic,ferrier2013phase,sticlet2014dynamical,dmytruk2016josephson,murani2016andreev} (see also the SM \cite{sm}), the susceptibility is the sum of three contributions, $\chi(\phi,\omega)\equiv\chi_{J}(\phi,\omega)+\chi_D(\phi,\omega)+\chi_{ND}(\phi,\omega)$, with $\chi_J(\phi,\omega)=\partial J_{s}/\partial\Phi$ and 
\begin{align}
\label{eq:suscept}
\chi_D&=\sum_n\frac{\omega}{\omega+i\gamma_D}\left(\frac{\partial\epsilon_n}{\partial\Phi}\right)^2\frac{\partial f(\epsilon_n)}{\partial\epsilon_n}\,,\nonumber\\
\!\!\chi_{ND}&=-\hbar\omega\sum_{n\neq m}\frac{|\langle m|\hat{I}_s|n\rangle|^2}{\epsilon_{nm}}\frac{f(\epsilon_n)-f(\epsilon_m)}{\epsilon_{nm}-\hbar\omega-i\hbar\gamma_{ND}}\,,
\end{align} 
corresponding to the Josephson, diagonal, and non-diagonal contributions, respectively \cite{NoteB}. Here, $J_s(\Phi)=-\sum_nf(\epsilon_n)\partial\epsilon_n/\partial\Phi$ is the Josephson current, with $\epsilon_n$ the single quasiparticle states, $\gamma_D$ and $\gamma_{ND}$ are the intra and inter-levels relaxation rates. Also, $f(\epsilon_n)$ is the equilibrium  occupation number of state $\epsilon_n$  which, in general, can depend on the constraints that we impose on the system. The susceptibility gives access to the level structure of the Andreev states, their phase dependence  and their population,  as well as the various relaxation rates associated with these levels. 
While the first (kinetic) and last (Kubo) terms have been analyzed in various setups, the second term is unique as it directly unravels the level structure around the zero energy (due to the derivative of the distribution function), and the time scales associated with these levels, as it has been shown experimentally in Ref~\onlinecite{dassonneville2013dissipation}. Moreover, the susceptibility is directly connected to the low-energy conductivity of the wire, as well as to the noise spectrum. The former is simply $\sigma(\phi,\omega)=(i/\omega)\chi(\phi,\omega)$, while the latter is found from the fluctuation-dissipation theorem as:
\begin{equation}
S(\phi,\omega)\simeq\hbar\coth{(\hbar\omega/2k_BT)}\chi''(\phi,\omega)\,,
\label{noise}
\end{equation}
which is dominated by  $\chi_D''(\phi,\omega)$ in the low frequency limit.
Furthermore,  $\chi_D$ being a low-frequency quantity, is non-invasive and thus more accessible experimentally~\cite{reulet1995dynamic,chiodi2011}.
This quantity  also  allows to distinguish between a genuine crossing  and an anti-crossing with a tiny gap. A priori, both would be very similar in transport. However, $\chi_D$ is able to distinguish a Landau-Zener process, which would happen for an anti-crossing, from a generic crossing due to the fact that both the frequency $\omega$ and the amplitude of the coupling can be independently controlled~\cite{trivedi1988mesoscopic,chiodi2011}.

In Fig.~\ref{Zeeman} (top), we plot the imaginary part of the susceptibility $\chi''(\phi,\omega)$  as a function of $\phi$ for several values of the Zeeman field $V_z$ in the topological regime. There are three main features associated with the response. First, the oscillation period of the susceptibility is $2\pi$, and not $4\pi$ as expected for the fractional Josephson effect~\cite{Nazarov-pikulin2011,san2012ac}. This is because our systems hosts $4$ instead of $2$ MBS and thus hybridization of these levels lifts the crossing at $E=0$ (see the SM for the full spectrum).  As shown previously~\cite{kitaev2001unpaired}, the overlap of the $\gamma_1$ and $\gamma_2$ MBS scales as $\sim\exp{(-l_{12}/\xi)}$, with $\xi$ the coherence length that, in the regime discussed in the work scales as $\xi\propto V_z$ \cite{KlinovajaPRB2011}.  Second, the signal evolves from a double-peak structure for Zeeman fields near the topological phase transition ($V_z\sim\Delta_{s}^w$), dominated by $\chi_D(\phi,\omega)$,  becoming a single peak around $\phi=\pi$ for larger Zeeman fields ($V_z\sim2\Delta_s^w$), where it is dominated by $\chi_{ND}(\phi,\omega)$ (for a comparison of the two contributions see SM). Third, as seen from the inset of Fig.~\ref{Zeeman} (top), the entire signal reduces as the Zeeman splitting is increased since the external MBSs overlap increases increasing  the splitting at the anti-crossing. We mention that according to Eq.~\eqref{noise}, the same features apply directly to the noise spectrum of the wire. This is one of our main results, namely that the dissipation is dominated by the diagonal term for a wide range of experimentally relevant parameters, previously disregarded in the literature. 

Let us now discuss  the parity dependence of the distribution functions $f(\epsilon_n)$. The many-body Hamiltonian can be written as $H_{w}=\sum_n\epsilon_n(d^\dagger_nd_n-1/2)$, with $d_n$ ($d_n^\dagger$) quantifying the annihilation (creation) of the Bogoliubov quasiparticle with energy $\epsilon_n$. The parity of the system is defined as $\tau=(-1)^{N}$, with $N=\sum_{n}d^\dagger_nd_n$, and the distribution carries changes if this is assumed to be conserved or not (thermodynamically). The thermal density matrix can be written as $\rho_\tau=P_\tau\exp(-\beta H_w)/Z_\tau$, with the projector  $P_\tau=[1+\tau(-1)^{N}]/2$, and $Z_\tau={\rm Tr}[P_\tau\exp(-\beta H_w)]$. Finally, the parity-dependent distribution function $f(\epsilon_n)\equiv f_\tau(\epsilon_n)={\rm Tr}[d_n^\dagger d_n\rho_\tau]$.  In the SM we summarize how precisely they depend on such a constraint and display the result for some simple case.  
For simplicity, we first discuss  the parity unrestricted case
and use a simplified model that only incorporates the low-energy subspace associated with four MBS.
   
\emph{Low-energy description--} The simplest Hamiltonian describing the four MBS low-energy spectrum reads:
\begin{align}
H_{M}=i\gamma_1(t_L\gamma_2+t'_L\gamma_3)+i(t_R\gamma_3+t'_{R}\gamma_2)\gamma_4+it_{LR}\gamma_2\gamma_3\,,
\end{align}   
with $t_{L(R)}$, $t'_{L(R)}$, and $t_{LR}\equiv t_{LR}(\phi)$ being the coupling between the first (last) two MBS, the first and third (second and fourth), and between the middle MBS, respectively. The couplings $t_{L,R}$ and $t'_{L,R}$ are assumed to depend on various external parameters, such as the chemical potential, Zeeman field, etc, but not on the phase $\phi$. On the other hand, $t_{LR}$ depends on the phase bias, and in the simplest models of tunneling, $t_{LR}(\phi)\propto\cos{(\phi/2)}$.  In the SM we consider more complex Hamiltonians with more coupling strengths. 
It is instructive to introduce the fermionic operators $c_A=(\gamma_3+i\gamma_2)/2$ ($c_A^\dagger=(\gamma_3-i\gamma_2)/2$) and $c_B=(\gamma_4+i\gamma_1)/2$ [$c_B^\dagger=(\gamma_4-i\gamma_1)/2$], so that the low-energy  Hilbert space is spanned by the states  $\{|00\rangle,c_A^\dagger|00\rangle,c_B^\dagger|00\rangle,c_A^\dagger c_B^\dagger|00\rangle\}$, with $|00\rangle$ being the vacuum with no electrons. The general state can be written as $|n_An_B\rangle$, with $n_A=0,1$ and $n_B=0,1$, with $\{|00\rangle,|11\rangle\}$ decoupled from the $\{|01\rangle,|10\rangle\}$ states due to parity conservation.
One can diagonalize the Hamiltonian in this basis to obtain both the eigenfunctions  and the many-body energies (see SM):
\begin{equation}
\!E_{\pm,\tau}(\phi)=\pm\sqrt{t^2_{LR}(\phi)+(t_L+\tau\,t_R)^2+(t'_L-\tau\,t'_R)^2}\,,
\end{equation}   
and the corresponding single particle energies $\epsilon_{1,2}=|E_{+,+}\pm E_{+,-}|$ (and the $-\epsilon_n$ partners), which can be inserted into the expression for the susceptibility. In Fig.~\ref{Zeeman} (bottom) we depict the many-body spectrum and the possible transitions with and without parity flips for some experimentally relevant parameters.  We mention that adding a term of the form $t_{14}\gamma_1\gamma_4$ (with $t_{14}$ the coupling strength between the outer MBS) pertains to the substitution $t_{LR}(\phi)\rightarrow t_{LR}(\phi)+\tau t_{14}$ in the above expression, and would change dramatically the spectrum as it becomes $4\pi$, instead of $2\pi$ periodic. 

Let us now consider both the cases when parity is unconstrained and constrained, respectively. The only difference in evaluating the susceptibilities comes from the distribution functions. In the unconstrained case, that is simply given by the FD function $f(\epsilon_n)=1/[1+\exp{(\beta\epsilon_n)}]$, while for the constrained case we find  $f_{1}(\epsilon_{1,2})=1/\{1+\exp{[\beta(\epsilon_1+\epsilon_2)]}\}$, and $f_{-1}(\epsilon_{1,2})=1/\{1+\exp{[\pm\beta(\epsilon_1-\epsilon_2)]}\}$ (see SM for details on the derivations).  The Josephson susceptibility for unconstrained and constrained parity, respectively, is given by  
\begin{align}
\chi_{J}&=\frac{\partial}{\partial\Phi}\sum_{n=1,2}\left[\tanh{\left(\frac{\beta\epsilon_n}{2}\right)}\frac{\partial\epsilon_{n}}{\partial\Phi}\right]\,,\\
\chi_{\tau,J}&=2\frac{\partial}{\partial\Phi}\left[\tanh{\left(\beta E_{+,\tau}\right)}\frac{\partial E_{+,\tau}}{\partial\Phi}\right]\,,
\end{align}
being independent of $\omega$, and where $\tau=\pm1$. The results are intuitive: the energies $E_{+,\tau}\equiv (\epsilon_1+\tau\epsilon_2)/2$ are nothing but the many-body energies for a given parity. Similarly, we evaluate the diagonal components in this low-energy subspace as
\begin{align}
\chi_{D}&=-\frac{\omega}{\omega+i\gamma_D}\sum_{n=1,2}\frac{\partial \tanh{(\beta\epsilon_{n}/2)}}{\partial\epsilon_{n}}\left(\frac{\partial\epsilon_{n}}{\partial\Phi}\right)^2\,,
\end{align}
for unconstrained parity, and
\begin{align}
\chi_{\tau,D}&=-\frac{2\omega}{\omega+i\gamma_D}\frac{\partial \tanh{(\beta E_{+,\tau})}}{\partial E_{+,\tau}}\left(\frac{\partial E_{+,\tau}}{\partial\Phi}\right)^2\,,
\end{align}
for constrained parity with $\tau=\pm1$. These expressions are our second main result. The diagonal susceptibility strongly depends on the constrained/unconstrained condition, and affects both the reactive  and dissipative  response  of the wire and thus should allow to probe whether parity is broken or not at the measured frequency.

Finally, the last contribution is due to the non-diagonal terms, or transitions between the levels and is given instead by: 
\begin{align}
\chi_{ND}&=-8\hbar\omega\left(\frac{\partial t_{LR}}{\partial\Phi}\right)^2\sum_{\tau=\pm1}[(t_L-\tau t_R)^2+(t'_L+\tau t'_R)^2]\nonumber\\
&\times\frac{f(\epsilon_{1})-f(\tau\epsilon_{2})}{(\epsilon_{1}-\tau\epsilon_{2})^3}\frac{\hbar\omega+i\hbar\gamma_{ND}}{(\epsilon_{1}-\tau\epsilon_{2})^2-(\hbar\omega+i\hbar\gamma_{ND})^2}\,,
\end{align}
for the unconstrained parity case, and
\begin{align}
\chi_{\tau,ND}&=\hbar\omega[(t_L-\tau t_R)^2+(t'_L+\tau t'_R)^2]\left(\frac{\partial t_{LR}}{\partial\Phi}\right)^2\nonumber\\
&\times\frac{\tanh{(\beta E_{+,\tau})}}{E_{+,\tau}^3}
\frac{\hbar\omega+i\hbar\gamma_{ND}}{4E_{+,\tau}^2-(\hbar\omega+i\hbar\gamma_{ND})^2}\,,
\end{align}
for the constrained parity case.

In Fig.~\ref{Parity} we plot the total imaginary part of susceptibility with and without the parity constraints as a function of $\phi$ and for different values of the frequency $\omega$. We chose the temperature $T$ such that $\chi_{D}''$ dominates for the unconstrained parity case at low frequencies, and by  $\chi_{ND}''$ in the constrained parity case in the entire frequency range. In the SM we discuss a larger range of parameter regime for $\chi_{D}''$ vs. $\chi_{ND}''$, and when the former dominates of the latter.  In general,  in order to  capture the full $\phi$-dependence both terms are important and need to be considered on equal footing, as in this work.    

The type of response (constrained vs. unconstrained parity) is dictated by the product $\omega\tau_p$, with $\tau_p$ being the parity lifetime $\tau_p$ (due, for example, to quasiparticle poisoning): it will correspond to $\chi(\phi,\omega)$ [$\chi_\tau(\phi,\omega)$] for $\omega\tau_p\ll1$ (for $\omega\tau_p\gg1$). However, in a typical experiment, for $\omega\tau_p\gg1$, the susceptibility will be a statistical mixture of the two parity states contributions $\chi_{\pm1}(\phi,\omega)$.  


\emph{Experimental realization --}
For the numerical calculations, the parameters we used correspond to $InSb$ nanowire and have been extracted from~\cite{mourik2012signatures} with the effective electron mass  $m^*=0.013m_e$, induced superconducting pairing $\Delta=0.25~m\text{eV}$ and spin-orbit hopping $\alpha_p=0.2~\text{eV}{\AA}$. We consider the nanowire of the length $2 l_W a =2~\mu\text{m}$ and take $l_W=40$ sites, which corresponds to the lattice constant $a=25~n\text{m}$. We take the hopping amplitude $t=\hbar^2/\left(2m^*a^2\right)=5~\text{meV}$ as an energy unit. In the tight-binding model, the spin-orbit hopping amplitude $\alpha = \alpha_p/2a = 0.4~\text{meV}$ and the chemical potential is tuned to $\mu = -10~\text{meV}$ . Following Ref.~\cite{dassonneville2013dissipation} and Ref.~\cite{albrecht2016transport}, we take the frequency to be $\omega = 200~\text{MHz}$  and the level lifetime $\gamma_D=\gamma_{ND}=0.1~\mu\text{s}^{-1}$, respectively.

\emph{Conclusions --} In this work, we studied the microwave response of a Josephson junction in a topological wire.
This response can be casted in three different contributions: one from the Josephson current, one stemming from Kubo response and involving transitions between the levels, and one diagonal term that requires finite temperature and which contains information about the levels coherences. Using a full numerical calculation, supplemented by a low-energy analytical solution, we have found that, at low frequencies and low temperatures, the dissipative response is dominated by the often neglected diagonal contribution of the current susceptibility.

\emph{Acknowledgments --} We would like to thank B. Dassonneville, M. Ferrier for discussions. M.~T. was supported by National Basic Research Program of China Grants No. 2011CBA00300 and No. 2011CBA00302, O.~D.  by the Swiss National Science Foundation and the NCCR QSIT, and R.~A. by the Spanish Ministry of Economy and Competitiveness through grant No. FIS2015-64654-P.


%

\clearpage
\widetext
\begin{center}
\huge{\textbf{Supplemental Material}}
\end{center}

\section{}
The topological SW connected to a SQUID as schematically depicted in Fig.~1 in the main text can host up to four MBS, depending on the system parameters. In this Supplemental Material (SM), we provide more details on the modeling of the system under consideration, on the derivation of the current susceptibility for both the general case and the parity conserving case, and finally we supply expressions for the susceptibility valid at low energy.

\section{Theoretical Model}

\begin{figure}[h] 
\centering
\includegraphics[width=0.45\linewidth]{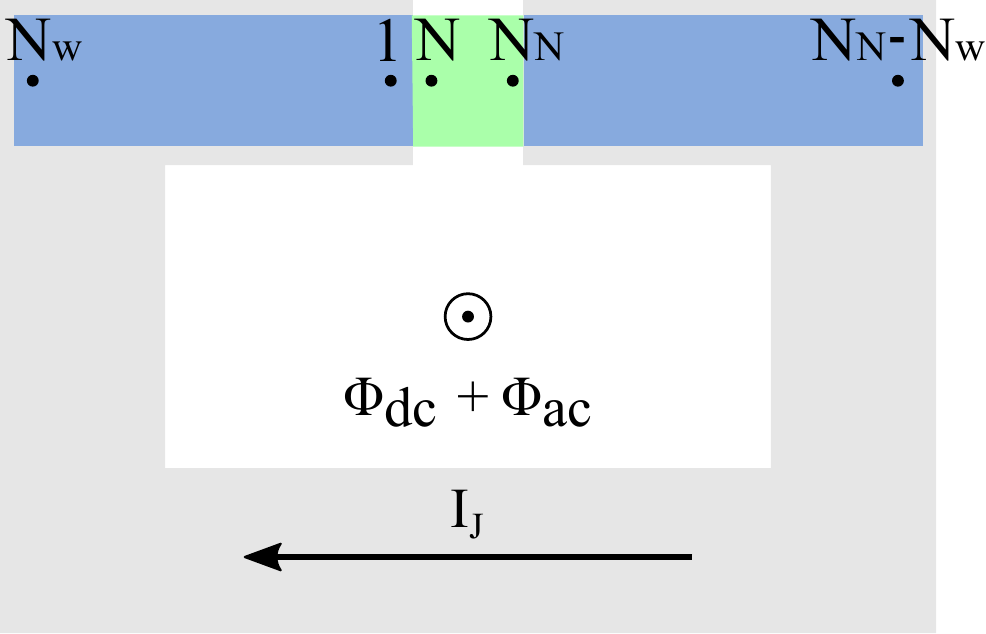}
\caption{Scheme of the setup. A nanowire (in blue) proximitized with bulk s-wave superconductor (in grey) of length and subject to both a dc and ac flux, $\Phi_{dc}$ and $\hat{\Phi}_{ac}$, respectively.  The bulk s-wave superconductor (in grey) is interrupted under the weak link (in green). } 
\label{fig:sketch} 
\end{figure}

We will first establish the conditions under which such a setup can be simply viewed as a wire under two superconductors carrying different superconducting phases (or a phase-biased wire). In order to model the system depicted in Fig.~1 in the main text, we consider a ring of total size $N$, (with the lattice spacing $a=1$),  with a weak link between sites $1$ and $N$. We assume the ring is  composed of two parts: a topological SW proximitized with a superconductor such as those analyzed in Ref.~\onlinecite{albrecht2016exponential1} which is separated into two pieces by the junction and an $s$-wave superconductor. We assume a ring geometry, with the following distribution of lengths (see Fig.~\ref{fig:sketch}):  the full wire length, $2l_w = 2(N_w - 1)a$, is defined for the sites $j$ fulfilling $1 \leq j \leq N_w$ and $N_N - N_w \leq j \leq N_N - 1$, the normal junction length $l_N = (N + 1 - N_N)a < 2l_w$ as $N_N \leq j \leq N + 1 \equiv 1$, while the rest is an $s$-wave superconductor of length $l_s = [N_N - 2(N_w+1)]a$, $N_w + 1 \leq j \leq N_N - N_w - 1$. The parts $1 \leq j \leq N_w$ and $N_N - N_w \leq j \leq N_N - 1$ are proximitized superconductors, that can become topological. 

In the following we describe in detail the  ac response of the topological wire in geometry  Fig.~1 in the main text. We first describe the setup in the presence of magnetic flux, bot dc and ac (the Hamiltonian pertains to discuss both on the same footing). The total Hamiltonian of the system can be written as  $H_{\rm sys} = H_{SW} + H_{T} + H_{S}$, with~\cite{rainis2013towards1}
\begin{align}
H_{sys}&=\sum_{j; \sigma,\sigma'=\uparrow,\downarrow}\Big[-t^*_{jj+1} c^\dag_{j+1,\sigma}\delta_{\sigma\sigma'}c_{j,\sigma'}-\mu c^\dag_{j,\sigma}\delta_{\sigma\sigma'}c_{j,\sigma'}+\Delta^*_{j} c^\dag_{j,\uparrow}c^\dag_{j,\downarrow},
-i\alpha_{jj+1}^* c^\dag_{j+1,\sigma}\sigma^y_{\sigma\sigma'}c_{j,\sigma'}-V_z c^\dag_{j,\sigma}\sigma^x_{\sigma\sigma'}c_{j,\sigma'}+h.c.\Big],
\label{eq:wirehamiltonian}
\end{align}
where $t_{jj+1}$ and $\alpha_{jj+1}$ are spin-orbit independent and dependent complex hopping matrix elements, respectively,  between the $j$ and $j+1$ sites. Here, $\Delta_j$ is the $s$-wave pairing at the position $j$, $\mu$ is the chemical potential,  $V_z$ is the Zeeman energy ($V_z=g\mu_BB/2$) and $\sigma_i$, with $i=x,y,z$ are the Pauli matrices. The tunneling matrix elements are as follows:
\begin{align}
t_{jj+1}&=\left\{
\begin{array}{cc}
te^{i\phi_{jj+1}} & 1 \leq j \leq N_N - 1\,,\\
t'e^{i\phi_{jj+1}} & N_N \leq j \leq N\,,
\end{array}
\right.
\end{align}
while for  $\alpha_{jj+1}$ we have:
\begin{align}
\alpha_{jj+1}&=\left\{
\begin{array}{cc}
\alpha e^{i\phi_{jj+1}} & 1 \leq j \leq N_w, N_N - N_w \leq j \leq N_N - 1\,,\\
\alpha' e^{i\phi_{jj+1}} & N_N \leq j \leq N\,,\\
0 & N_w + 1 \leq j \leq N_N - N_w - 1\,,
\end{array}
\right.
\end{align}

We assumed that the tunneling matrix elements $t$ are the same but in the region that does not contain a superconductor (the normal region). While such an assumption can look simplistic, it covers the relevant physics. The phases $\phi_{jj+1}$ are given by the usual expression:  
\begin{align}
\Phi_{j,j+1}(t)&=\frac{e}{\hbar}\int_{j}^{j+1}dx A(x)\equiv\frac{e\Phi_{tot}(t)}{\hbar N}\,,
\end{align}  
where $A(x)$ in the vector potential along the loop. For the superconducting pairing we can write instead:
\begin{align}
\Delta_{j}&=\left\{
\begin{array}{cc}
\Delta e^{i\phi_{j}} & N_w + 1 \leq j \leq N_N - N_w - 1\\
\Delta_w e^{i\phi_{j}} & 1 \leq j \leq N_w, N_N - N_w \leq j \leq N_N - 1\,,\\
0 & N_N \leq j \leq N\,,
\end{array}
\right.
\end{align}
where the phase $\phi_j$ needs to be found self-consistently from the conditions imposed on the top $s$-wave superconductor. If the top superconductor is disconnected, i.e. it does not allow for a super-current flow through it, the phase $\phi_j$ can be established from the following condition:
\begin{align}
J_s=\frac{2e}{m}|\psi|^2(\hbar\nabla\phi-2eA)\equiv0\,,
\end{align}
which leads to the result obtained  in the Main text:
\begin{equation}
\phi(x)=\frac{2e \Phi_{tot}(t)}{\hbar}\frac{x}{L}\,.
\end{equation}

\begin{figure}[t] 
\centering
\includegraphics[width=0.9\linewidth]{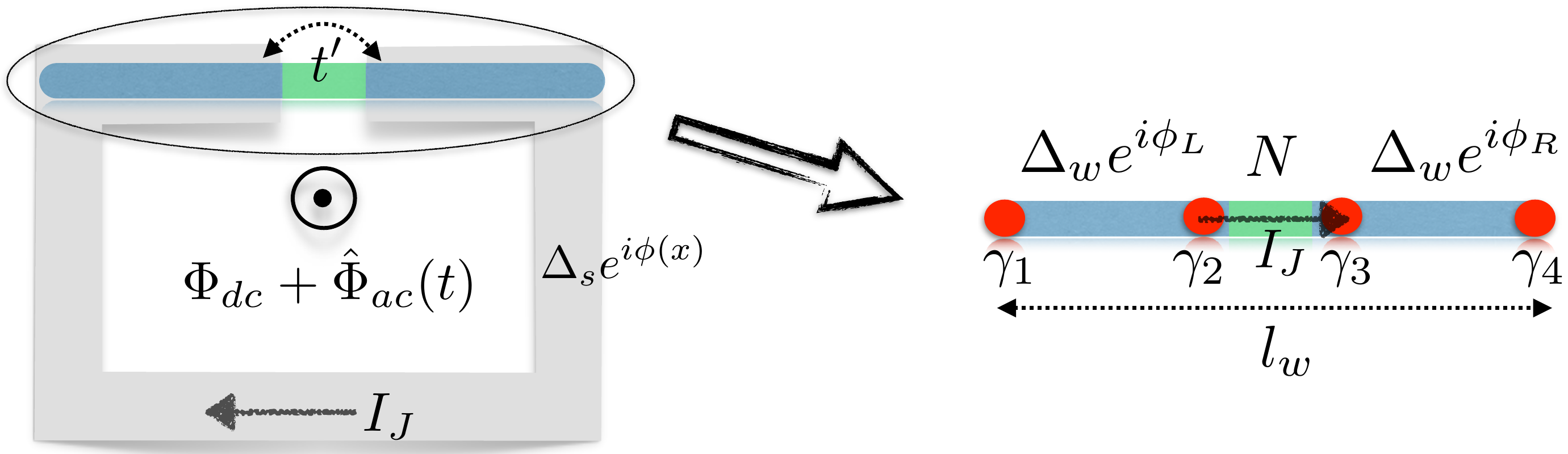}
\caption{Scheme of the setup. Left:  A nanowire (in blue) proximitized with bulk s-wave superconductor (in grey) of length $L=Na$  and subject to both a dc and ac flux, $\Phi_{dc}$ and $\hat{\Phi}_{ac}(t)$, respectively.  The bulk s-wave superconductor (in grey) is interrupted under the weak link (in green). Right: the resulting effective model pertaining to a SNS system, with a phase bias over the weak link. Black dots indicate four Majorana fermions $\gamma_1 - \gamma_4$.} 
\label{fig:schemeSM} 
\end{figure}

Note that for the underneath superconductor we assumed the continuum description, so that in the lattice model describing the wire we need to substitute $x\rightarrow j$. The above condition on the phase holds true only approximatively as a finite super-current changes the effective flux threaded through the ring. However, we adopt the usual (experimental) assumption that both the  kinetic inductance of the $s$-wave superconductor and the geometrical inductance of the total ring (including the wire) is much smaller than that of the wire, and thus these two are assumed not to cause any feed-back on the applied bare flux. We see that for a large ring (which is mostly composed of the superconductor), the phase influence is most important on the superconducting part, which can be of the order of the full flux $\Phi$ for short wires. Since we assume the large superconductor ``inert", namely that its states are not influenced by the presence of the wire and the flux (under the above assumptions), we can assume that on the left and on the right of the wire they act as infinite leads with phases $\phi_L(t)=0$, and $\phi_R(t)\equiv\phi(t)=2e\Phi_{tot}(t)/\hbar$, respectively. The resulting Hamiltonian can then be written as $H = H_L+H_R+H_w$, where $H_{L,R}$ describe conventional $s$-wave superconductors with pairings $\Delta_{s}=\Delta(\delta_{s,L}+\delta_{s,R}e^{i\phi(t)})$, where $H_{s=L,R,w}$ reads:
\begin{align}
H_{s}&=\sum_{s=L,N,R,j\in s; \sigma,\sigma'=\uparrow,\downarrow}\Big[-t_s c^\dag_{j+1,\sigma}\delta_{\sigma\sigma'}c_{j,\sigma'}+\Delta_s^* c^\dag_{j,\uparrow}c^\dag_{j,\downarrow}
-i\alpha_s c^\dag_{j+1,\sigma}\sigma^y_{\sigma\sigma'}c_{j,\sigma'}-\mu c^\dag_{j,\sigma}\delta_{\sigma\sigma'}c_{j,\sigma'}-V_z c^\dag_{j,\sigma}\sigma^x_{\sigma\sigma'}c_{j,\sigma'}+h.c.\Big]\,,
\end{align} 
where $\Delta_{L,R}=\Delta_w(\delta_{s,L}+\delta_{s,R}e^{i\phi(t)})$, and $\Delta_N=0$ (in the central normal part). That emulates the general setup $SNS$ described in many previous works. The total flux can be decomposed into a {\it dc} and {\it ac} component, respectively: $\phi(t)=\phi+\delta\phi(t)$, with the latter serving as to probe the spectral properties of the wire.  
One can find the single particle excitations of this many-body Hamiltonian by writing it in the BdG basis as $H=\vec{c}^\dagger H_{BdG}\vec{c}$, with 
\begin{equation}
\vec{c}^\dagger\equiv(c^\dagger_{1\uparrow}, c^\dagger_{1\downarrow},\dots c^\dagger_{N\uparrow}, c^\dagger_{N\downarrow},c_{1\uparrow}, c_{1\downarrow},\dots c_{N\uparrow}, c_{N\downarrow})
\end{equation}
and $H_{BdG}$ being a $4N\times4N$ matrix describing the single particle spectrum. The time-dependent perturbation preserves this form of the Hamiltonian and thus can be described in the single particle language as long as the mean-field Hamiltonian only is considered, so that $H_{BdG}\rightarrow H_{BdG}(t)$.  

We mention that for $V_z^s > \sqrt{(\mu_s+2t)^2 + (\Delta^s_w)^2}$, there are four Majorana fermions present in the ring: $\gamma_1$, $\gamma_4$ localized at the ends of the spin-orbit coupled nanowire, and $\gamma_2$, $\gamma_3$ localized on both sides of the weak link (see Fig.\ref{fig:schemeSM}).  Otherwise, no Majorana fermions emerges and the system is in the topologically trivial phase, or the four Majorana fermions fuse and they move into the (wire) bulk spectrum.  

\section{Derivation of the finite-frequency response}

Next we focus on the Hamiltonian $H_{BdG}(t)$ in the presence of both {\it dc} and {\it ac} magnetic fluxes (we will write $H(t)$ from now on to simplify the notations). Let us decompose the Hamiltonian $H(t)$ into the static and the time-dependent contribution (in leading order in the {\it ac} flux $\delta\Phi(t)$). We get:
\begin{align}
H(t)&=H_0+V(t)\,,\\
V(t)&=-\hat{I}_s\delta\Phi(t) = \frac{\partial H_0}{\partial\Phi}\delta\Phi(t) \,,
\end{align}
where $H_0$ is the Hamiltonian in the absence of the driving, and $\hat{I}_s$ is the current operator in the absence of the drive (for a nice derivation of this expression see Ref.~\onlinecite{vayrynen2015microwave1}, where they show explicitly that in the low energy limit, only tunneling of pairs is responsible for the current and considering only the proximity effect is sufficient to calculate all the transport quantities, i.e. no need to consider the superconducting leads that provide it). Next we need to for the response of the wire to the {\it ac} perturbation. In Ref.~\onlinecite{trivedi1988mesoscopic1}, it was established the general out-of-equilibrium equation for the reduced density matrix of a {\it normal} ring in the presence of environment, and consequently in the presence of relaxations. They assume the weak coupling limit to the environment, for which they found:
\begin{equation}
\frac{\partial \rho(t)}{\partial t}+\dfrac{i}{\hbar}[H(t),\rho(t)]=-\Gamma[\rho(t)-\rho_{qe}(t)]\,,
\end{equation}
with $\rho(t)$ being the reduced density matrix of the system (after tracing over the environment), $H(t)$ being the total (time-dependent) Hamiltonian, $\Gamma$ is the reduced relaxation tensor (that accounts for both the diagonal and off-diagonal relaxations), and 
\begin{equation}
\rho_{qe}(t)=\frac{1}{1+e^{H(t)/k_B T}}\,,
\end{equation}
being the instantaneous quasiequilibrium density matrix of the system (time-dependent) at the single-particle level. However, the qusiequilibrium density matrix could be a general and not necessary the one above. We will discuss that in the next section when addressing the case when the parity of the system is conserved. We note again that $H(t)$ represents the single-particle Hamiltonian, with the condition that in the absence of the drive $H_0|m\rangle=\epsilon_m|m\rangle$, with $\epsilon_m$ and $|m\rangle$ being the quasiparticle energies and eigenvectors, respectively in the absence of the drive.  We are left with evaluating the full density matrix of the system in leading order in the perturbation. First, let us find the quasi-equilibrium component, $\rho_{qe}(t)$.  We get:
\begin{align}
\langle n|\rho_{qe}(t)|m\rangle&=f(\epsilon_n)\delta_{nm}+\frac{f(\epsilon_n)-f(\epsilon_m)}{\epsilon_n-\epsilon_m}\langle n|V(t)|m\rangle\,,\\
\end{align}
where $\{|n\rangle\}$ and $\{\epsilon_n\}$  are eigenvectors and eigenstates of the bare Hamiltonian in the absence of the driving, respectively, and $\rho_0|m\rangle=f(\epsilon_m)|m\rangle$. These are in fact the single particle states that build up the Slater determinant that describes the many-body states. However, in the case of superconducting systems, $V(t)$ can allow for a change of particle number by Cooper pairs, thus $|n\rangle$ and $|m\rangle$ can describe states with $\pm2$ electrons. We need to be aware of such a feature later in the calculation. Note that the full density matrix can be written as $\rho(t)=\rho_0+\delta\rho(t)$, with $\rho_0$ and $\delta\rho$ corresponding to the density matrix in the absence of the perturbation, and the deviation from that, respectively. In this work, we consider monochromatic drives, of the sort $\delta\Phi(t)=\delta\Phi(\omega)\exp{(-i\omega t)}$, with $\omega$ the driving frequency,  which allows us to write:
\begin{align}
&\rho(t)=\rho_0+\delta\rho(\omega)e^{-i\omega t}\,,\\
&V(t)=V(\omega)e^{-i\omega t}\,,
\end{align}
which in turn gives rise to the following equation for the density matrix deviation:
\begin{equation}
-\hbar\omega\delta\rho(\omega)+[H_0,\delta\rho(\omega)]+[V(\omega),\rho_0]=i\hbar\Gamma[\delta\rho(\omega)-\delta\rho_{qe}(\omega)]\,.
\end{equation}
With that, we can readily calculate the matrix elements of the time-dependent density matrix in the bare basis:
\begin{align}
\langle n|\delta\rho(\omega)|m\rangle&=\frac{\epsilon_n-\epsilon_m-i\hbar\gamma_{nm}}{\epsilon_n-\epsilon_m-\hbar\omega-i\hbar\gamma_{nm}}\frac{f(\epsilon_n)-f(\epsilon_m)}{\epsilon_n-\epsilon_m}\langle n|V(\omega)|m\rangle\nonumber\\
&=-\frac{\epsilon_n-\epsilon_m-i\hbar\gamma_{nm}}{\epsilon_n-\epsilon_m-\hbar\omega-i\hbar\gamma_{nm}}\frac{f(\epsilon_n)-f(\epsilon_m)}{\epsilon_n-\epsilon_m}\langle n|\hat{I}_s|m\rangle\,\delta\Phi(\omega)\,,\\
\langle n|\delta\rho(\omega)|n\rangle&=-\frac{i\gamma_{nn}}{\omega+i\gamma_{nn}}\frac{\partial f(\epsilon_n)}{\partial\epsilon_n}\langle n|\hat{I}_s|n\rangle\,\delta\Phi(\omega)
\end{align}
with $\gamma_{nm}\equiv[\Gamma]_{nm}$ being the $nm$ component of the relaxation tensor. We are now in position to calculate the average (time-dependent) current flowing through the system, which is given as $\langle \hat{I}_{s}(t)\rangle\equiv{\rm Tr}[\hat{I}_{s}(t)\rho(t)]$, with 
\begin{equation}
\hat{I}_{s}(t)=-\frac{\partial H(t)}{\partial\Phi(t)}\,. 
\end{equation} 
We are interested in the linear response regime, so we separate the current operator into a bare and a linear contribution in the drive:
\begin{align}
\hat{I}_s(t)&=\hat{I}_s+\delta \hat{I}_s(t)\,,\\
\delta \hat{I}_s(t)&=-\delta\Phi(t)\frac{\partial^2H_0}{\partial\Phi^2}\,,
\end{align}
this last term being known as the diamagnetic current. Putting everything together, we get for the average current:
\begin{align}
\langle \hat{I}_s(t)\rangle={\rm Tr}[\hat{I}_s\rho_0]+{\rm Tr}[\hat{I}_s\delta\rho(t)]+{\rm Tr}[\delta\hat{I}_s(t)\rho_0]\,.
\end{align}
We are interested in the change in the average current induced by the perturbation, thus we define $\delta\langle\hat{I}_s\rangle\equiv\langle \hat{I}_s(t)\rangle-{\rm Tr}[\hat{I}_s\rho_0]={\rm Tr}[\hat{I}_s\delta\rho(t)]+{\rm Tr}[\delta\hat{I}_s(t)\rho_0]$, and consequently on the 
susceptibility:
\begin{equation}
\chi(\Phi,\omega)=\frac{\delta\langle\hat{I}_s\rangle}{\delta\Phi(\omega)}\,,
\end{equation}
which quantifies the linear response of the wire. That is the final quantity we are after. We continue by writing in detail the induced charge current $\delta\langle\hat{I}_s\rangle$ using the matrix elements for $\delta\rho(t)$ found above:
\begin{align}
\chi(\Phi,\omega)&=-\sum_nf(\epsilon_n)\langle n|\frac{\partial^2H_0}{\partial\Phi^2}|n\rangle-\sum_{n\neq m}|\langle m|\hat{I}_s|n\rangle|^2\frac{\epsilon_n-\epsilon_m-i\hbar\gamma_{nm}}{\epsilon_n-\epsilon_m-\hbar\omega-i\hbar\gamma_{nm}}\frac{f(\epsilon_n)-f(\epsilon_m)}{\epsilon_n-\epsilon_m}\nonumber\\
&-\sum_n(\langle n|\hat{I}_s|n\rangle)^2 \frac{i\gamma_{nn}}{\omega+i\gamma_{nn}}\frac{\partial f(\epsilon_n)}{\partial \epsilon_n}\,.
\end{align}
It was shown that there are a couple of sum rules that help reducing more the above expression, and they read as follows \cite{trivedi1988mesoscopic1,dassonneville2013dissipation1}, and we adapt those situations to our superconducting system. The BdG Hamiltonian can be diagonalized as
\begin{align}
H_0=\sum_{p=1}^{4N}\epsilon_p(\Phi)|p(\Phi)\rangle\langle p(\Phi)|\,,
\end{align}
with $\epsilon_p(\Phi)$ the flux-dependent single particle energies,  and  $|p(\Phi)\rangle$ the single-particle (flux-dependent) wavefunctions.  We see that this Hamiltonian results in particle-hole symmetric eigenvalues which we need to account for when evaluating the susceptibility. Now let us add a small change to the {\it dc} flux, $\Phi\rightarrow\Phi+\delta\Phi$, with  $\delta\Phi$ a small deviation. The single particle energy can be written as:
\begin{equation}
\delta\epsilon_p(\Phi)\equiv\epsilon_p(\Phi+\delta\Phi)-\epsilon_p(\Phi)=\delta\Phi\frac{\partial\epsilon_p(\Phi)}{\partial\Phi}+\frac{(\delta\Phi)^2}{2}\frac{\partial^2\epsilon_p(\Phi)}{\partial\Phi^2}+\dots\,.
\end{equation}  
Next we can evaluate the deviation of the energy by using the perturbation theory on the modified Hamiltonian. The change in the Hamiltonian, in second order, caused by a small variation of the flux reads:
\begin{align}
\delta H&=\sum_j\left(-i\dfrac{2e}{\hbar}\delta\Phi-\left(\dfrac{2e}{\hbar}\right)^2\frac{(\delta\Phi)^2}{2}\right)\Delta e^{-i\phi}c_{j,\uparrow}^\dagger c_{j,\downarrow}^\dagger+{\rm H. c.}\equiv-\hat{I}_s\delta\Phi+\frac{1}{2}\frac{\partial^2H_0}{\partial\Phi^2}(\delta\Phi)^2.
\end{align}

We can then find the change in energy of a state $|n\rangle$ due to this perturbation:
\begin{align}
\delta \epsilon_p(\Phi)&=\langle p|\delta H|p\rangle+\sum_{m\neq p}\frac{|\langle m|\delta H|p\rangle|^2}{\epsilon_p-\epsilon_m}=-\langle p|\hat{I}_s|p\rangle\delta\Phi+\frac{1}{2}\langle p|\frac{\partial^2H_0}{\partial\Phi^2}|p\rangle(\delta\Phi)^2+\sum_{m\neq p}\frac{|\langle p|\hat{I}_s|m\rangle|^2}{\epsilon_p-\epsilon_m}(\delta\Phi)^2\,.
\end{align}
With that, we can identify the following identities:
\begin{align}
&\langle p|\hat{I}_s|p\rangle=-\frac{\partial\epsilon_{p}}{\partial\Phi}\,,\\
&\langle p|\frac{\partial^2H_0}{\partial\Phi^2}|p\rangle+2\sum_{m\neq p}\frac{|\langle p|\hat{I}_s|m\rangle|^2}{\epsilon_p-\epsilon_m}=\frac{\partial^2\epsilon_p}{\partial\Phi^2}\,.
\end{align}
which, when inserted into the expression for the susceptibility leads to:
\begin{align}
\chi(\Phi,\omega)&=\underbrace{\frac{\partial I_{J}}{\partial\Phi}}_{\chi_{J}}+\underbrace{\sum_n\frac{\omega}{\omega+i\gamma_{nn}}\left(\frac{\partial\epsilon_n}{\partial\Phi}\right)^2\frac{\partial f(\epsilon_n)}{\partial\epsilon_n}}_{\chi_D}-\underbrace{\hbar\omega\sum_{n\neq m}\frac{|\langle m|\hat{I}_s|n\rangle|^2}{\epsilon_n-\epsilon_m}\frac{f(\epsilon_n)-f(\epsilon_m)}{\epsilon_n-\epsilon_m-\hbar\omega-i\hbar\gamma_{nm}}}_{\chi_{ND}}\,.
\label{susc}
\end{align}
with 
\begin{equation}
I_{J}(\Phi)=-\sum_nf(\epsilon_n)\frac{\partial\epsilon_n}{\partial\Phi}\,,
\end{equation} 
being the super-current flowing in the presence of the static flux. The susceptibility can thus be decoupled into three parts: the Josephson ($\chi_J$), the diagonal ($\chi_D$), and the non-diagonal ($\chi_{ND}$, or Kubo) contributions, respectively. The first and the last ones have been the subject of several previous works, but the second term have been missed from most of the calculations. We mention that the sum is over all the states, both occupied and unoccupied.

\begin{figure}[t] 
\centering
\includegraphics[width=0.99\linewidth]{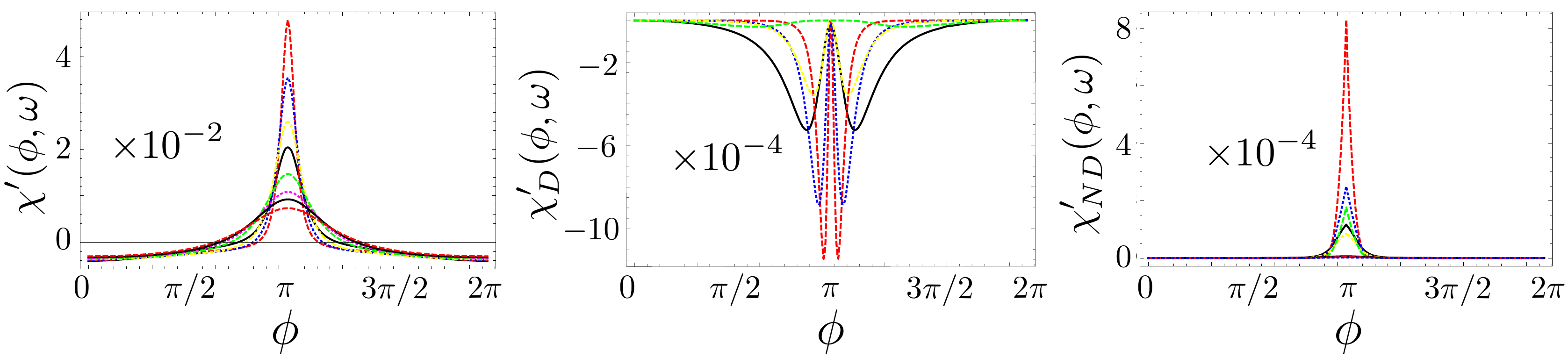}\
\caption{The total (left), diagonal (middle), and the non-diagonal  (right) contributions to the real part of the susceptibility  as a function of $\phi$ for different values of the Zeeman field $V_z$.  The black, red, blue, brown,  yellow, and green  lines correspond to $V_Z=1.2, 1.3, 1.4, 1.5$, $1.6$, and $1.7$ ($\times\,\Delta_s^w$), respectively.   The topological transition takes place at $V_z=1$. We expressed all energies in terms of the hopping $t$, with $\Delta_s^w=0.05$, $\alpha=0.08$, $\omega=1.6\times10^{-4}$, $T=0.8\,\omega$, $\gamma=10^{-8}$.} 
\label{ReZeeman} 
\end{figure}
In Fig~\ref{ReZeeman} we plot the real part of the total (left), the diagonal (middle), and the non-diagonal (right) susceptibility, respectively, as a function of $\phi$ for different values of the Zeeman field  $V_z$. We do not plot the Josephson susceptibility separately as it can be seen (from the left plot) that this practically dominates the reactive (real part) response. 
\begin{figure}[t] 
\centering
\includegraphics[width=0.9\linewidth]{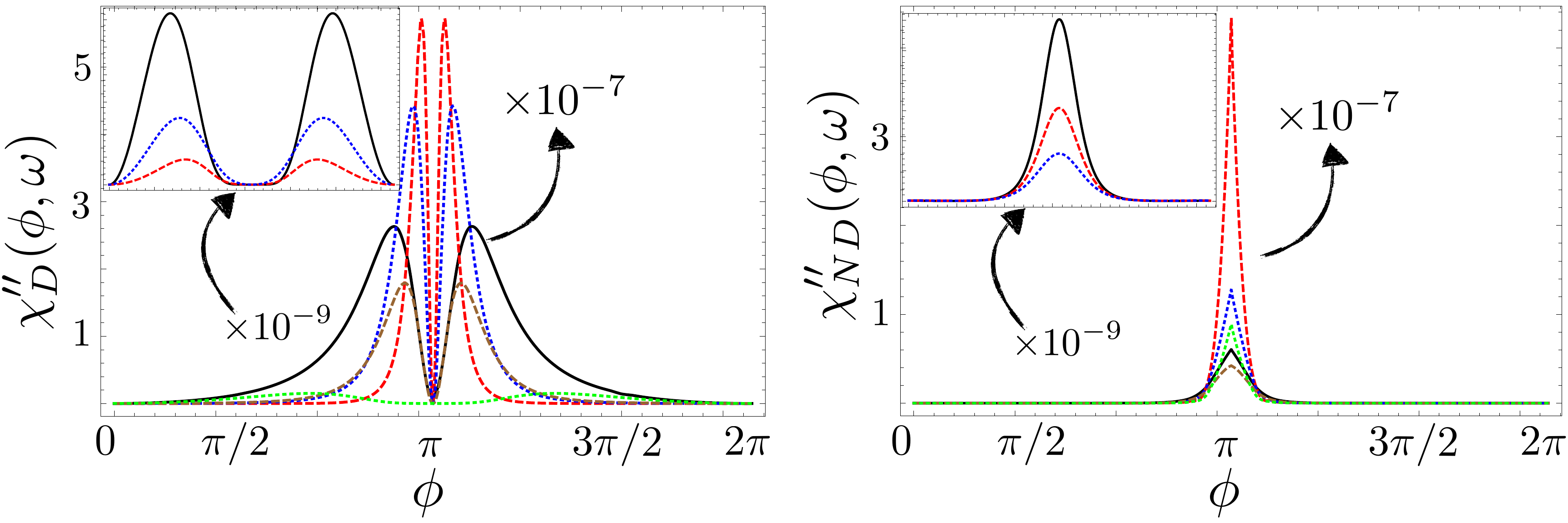}\
\caption{Left (Right): The diagonal (non-diagonal) contribution to the imaginary part of the susceptibility $\chi_D''(\phi,\omega)$ ($\chi_{ND}''(\phi,\omega)$) as a function of $\phi$ for different values of the Zeeman field $V_z$.  The black, red, blue, brown, and  green  lines correspond to $V_Z=1.2, 1.3, 1.4, 1.5$, and  $1.6$ ($\times\,\Delta_s^w$), respectively  (Inset: The  black, red, blue lines correspond to $V_Z=1.7, 2$, and  $2.3$, respectively).  The topological transition takes place at $V_z=1$. We expressed all energies in terms of the hopping $t$, with $\Delta_s^w=0.05$, $\alpha=0.08$, $\omega=1.6\times10^{-4}$, $T=0.8\,\omega$, $\gamma=10^{-8}$.} 
\label{ImZeeman} 
\end{figure} 
In Fig.~\ref{ImZeeman} we show separately the  diagonal (left) and the non-diagonal (right) contributions to the imaginary part of the susceptibility. The dissipative response is dominated by the diagonal contribution at low Zeeman fields and by the non-diagonal part at higher fields. Nevertheless, the response is in general reduced at larger fields.   

\section{Parity conservation effects on the currents and susceptibility}

In the previous section we analyzed the resulting expression for the susceptibility indifferent of any parity constraints. However, for our discussion here, parity plays a major role, and in superconductor is can be a good quantum number that can survive (in principle) to thermodynamic averaging. Let us here develop a parity dependent susceptibility, for all components. Note that the Hamiltonian can be written as:
\begin{equation}
H_{0}=\sum_n\epsilon_n(d_n^\dagger d_n-1/2)\,,
\end{equation}
and the parity of the system  defined as $\hat{\tau}=(-1)^{N}$, with $N=\sum_{n}d_n^\dagger d_n$. The parity-dependent partition function is then given by:

\begin{align}
Z_{\tau}&=\frac{1}{2}{\rm Tr}[(1+\tau(-1)^{N})e^{-\beta H_{0}}]=\frac{1}{2}\left({\rm Tr}[e^{-\beta H_{0}}]+\tau{\rm Tr}[(-1)^{N}e^{-\beta H_{0}}]\right)\nonumber\\
&=\frac{1}{2}\left[\prod_j(e^{\beta\epsilon_j/2}+e^{-\beta\epsilon_j/2})+\tau\prod_j(e^{\beta\epsilon_j/2}-e^{-\beta\epsilon_j/2})\right]=\frac{Z_0}{2}\left[1+\tau\underbrace{\prod_j\tanh{(\beta\epsilon_j/2)}}_{A}\right]\,,
\end{align} 
with $Z_0=\prod_j(e^{\beta\epsilon_j/2}+e^{-\beta\epsilon_j/2})$ the partition function without the parity constraint, $\beta=1/k_B T$ and $\tanh(x)=(e^x-e^{-x})/(e^x+e^{-x})$. 
The resulting free energy is given by $F_{\tau}=-\beta^{-1}{\rm ln}Z_{\tau}\equiv F_0+\tau F_a$, with:
\begin{align}
F_0&=-\beta^{-1}\left\{\ln (Z_0/2)+\dfrac{1}{2}\ln[(1-A^2)]\right\}\,,\\
F_a&=-\beta^{-1}\dfrac{1}{2}\ln\left(\frac{1+A}{1-A}\right)\,.
\end{align}
With these expressions, the Josephson current can be easily computed as:

\begin{align}
I_{J,\tau}&=\frac{2e}{\hbar}\frac{\partial F_{\tau}(\phi)}{\partial\phi}=\frac{\partial F_{\tau}(\Phi)}{\partial\Phi}=I_{J,0}+\tau I_{J,a}\,,\\
I_{J,0}&=\frac{\partial F_0}{\partial\Phi}=-\beta^{-1}\Big[\dfrac{1}{Z_0}\frac{\partial Z_0}{\partial\Phi} -  \frac{A}{1-A^2}\frac{\partial A}{\partial\Phi}\Big] =\sum_{j}\left\{f(\epsilon_j)-1/2+\frac{A^2}{1-A^2}\frac{1}{\sinh{(\beta\epsilon_j)}}\right\}\frac{\partial \epsilon_j}{\partial\Phi} \,,\\
I_{J,a}&=\frac{\partial F_a}{\partial\Phi} = -\beta^{-1}\frac{1}{1-A^2}\frac{\partial A}{\partial\Phi} = -\frac{A}{1-A^2}\sum_{j}\frac{1}{\sinh{(\beta\epsilon_j)}}\frac{\partial \epsilon_j}{\partial\Phi}\,,
\end{align}

which different from the usual current in the absence of parity constraint.  Let us also calculate the distribution function  for the quasiparticles. First, note that the density matrix can be written as
\begin{equation}
\rho_\tau=\frac{[1+\tau(-1)^{N}]e^{-\beta H_0}}{2Z_\tau}\,,
\end{equation}
so that

\begin{align}
f_\tau(\epsilon_j)&={\rm Tr}[d^\dagger_j d_j\rho_\tau]=-\frac{1}{\beta Z_\tau}\frac{\partial Z_\tau}{\partial\epsilon_j}+\frac{1}{2}=-\frac{Z_0}{4Z_\tau}\left[\coth{(\beta\epsilon_j/2)}+\tau\frac{4}{\exp{(\beta\epsilon_j)}+\exp{(-\beta\epsilon_j)}}A\right]+\frac{1}{2}\,,
\end{align}
which again differs from just the Fermi-Dirac distribution. Let us see how the distribution function looks like for various cases. First, we assume just one level system. In this case, we obtain that $f_{1(-1)}(\epsilon_1)=0(1)$ for all temperatures, as expected. This case correspond to a Majorana junction (just two Majoranas). That distribution can be used to derive the full parity-dependent susceptibility response. For the case of two levels, $\epsilon_j$ with $j=1,2$, we obtain the following distribution functions:
\begin{align}
f_{\tau}(\epsilon_{1})&=\frac{1}{1+e^{\beta[\epsilon_1+(-1)^\tau\epsilon_2]}}\,,\\
f_{\tau}(\epsilon_{2})&=\frac{1}{1+e^{\beta[\epsilon_2+(-1)^\tau\epsilon_1]}}\,.
\end{align}
Finally, we also give the result for three levels $\epsilon_j$ with $j=1,2,3$, namely we obtain:
\begin{align}
f_{\tau}(\epsilon_{1})&=\frac{1}{1+e^{\beta\epsilon_1}\cosh{[\beta(\epsilon_2+(-1)^\tau\epsilon_3)/2]}/\cosh{[\beta(\epsilon_2-(-1)^\tau\epsilon_3)/2]}}\,,\\
f_{\tau}(\epsilon_{2})&=\frac{1}{1+e^{\beta\epsilon_2}\cosh{[\beta(\epsilon_1+(-1)^\tau\epsilon_3)/2]}/\cosh{[\beta(\epsilon_1-(-1)^\tau\epsilon_3)/2]}}\,,\\
f_{\tau}(\epsilon_{3})&=\frac{1}{1+e^{\beta\epsilon_3}\cosh{[\beta(\epsilon_1+(-1)^\tau\epsilon_2)/2]}/\cosh{[\beta(\epsilon_1-(-1)^\tau\epsilon_2)/2]}}\,,
\end{align}
which clearly shows deviations from the usual FD distribution, 
\begin{equation}
f_{FD}(\epsilon_j)=\frac{1}{1+e^{\beta\epsilon_j}}\,.
\end{equation}
That simply means that the susceptibility in Eq.~\eqref{susc}, $\chi(\omega,\Phi)$ should be switched to $\chi_\tau(\omega,\Phi)$, namely that it does depend on the parity $\tau$ of the superconducting system if parity is conserved.

\section{Effective low energy model}

\subsection{Two Majoranas}
The first approximation is to consider only the lowest energy state, which is the mixture of the end modes Majoranas (2 in total). The subspace spanned by the 2 Majoranas:
\begin{equation}
H_{M}=i\epsilon(\phi)\gamma_2\gamma_3 = -\epsilon(\phi)(2c_A^\dagger c_A-1)\,,
\end{equation} 
where we note that for a two tunnel-coupled Majoranas $\epsilon=t_{LR}\cos{(\phi/2)}$, with $t_{LR}$ being the phase dependent coupling strength.

In the typical  tunnel junction coupling two Majorana bound states,  if parity is not conserved, the susceptibility reads:
\begin{align}
\chi(\omega,\phi)&=\chi_J+\chi_D\,,\\
\chi_{J}&=-\frac{\partial}{\partial\Phi}\left[\left(2f(\epsilon)-1\right)\frac{\partial\epsilon}{\partial\Phi}\right]\,,\\
\chi_D&=-\frac{2i\omega}{\gamma-i\omega}\frac{\partial f(\epsilon)}{\partial\epsilon}\left(\frac{\partial\epsilon}{\partial\Phi}\right)^2\,,
\end{align} 
with $\chi_{ND}\equiv0$ as there are no transitions possible that conserve the parity. 
 
We see that there is an imaginary (dissipative) component coming from the diagonal component, and which reads:
\begin{align}
\chi''_D(\omega,\phi)&=\left(\dfrac{2e}{\hbar}\right)^2\frac{t_{LR}^2\omega\gamma}{2(\gamma^2+\omega^2)}\frac{\beta e^{\beta\epsilon}}{(1+e^{\beta\epsilon})^2}\sin^2{(\phi/2)}\,.
\end{align}
However, the results change dramatically if parity is assumed as a constraint. The susceptibility is given simply by:

\begin{equation}
\chi_\tau(\omega,\phi)\equiv\chi_{\tau,J}(\omega,\phi)=\tau\frac{\partial^2\epsilon}{\partial\Phi^2}=-\frac{\tau}{4}\left(\dfrac{2e}{\hbar}\right)^2 t_{LR}\cos{(\phi/2)}\,,
\end{equation}

since the distribution functions are energy independent for the case of one energy level, as discussed above and thus all other terms vanish. Thus, as expected, no dissipative component exists in this situation. 

\subsection{Four Majoranas}

While the expression for the susceptibility discussed in the previous sections contains in principle all the energy levels, the the strongest response comes in fact from the low-energy levels close to the Fermi level, and in the following we assume only those. The lowest levels are the in-gap Andreev states hosted by the normal region, which contain also the Majorana (end) modes.

\subsubsection{Hamiltonian and spectrum for four Majoranas}

The second approximation is to consider only the lowest energy states, which are the mixture of the end modes Majoranas (4 in total). The most general Hamiltonian in the subspace spanned by the 4 Majoranas reads:
\begin{equation}
H_{M}=i\gamma_1(t_L\gamma_2+t_L'\gamma_{3})+i(t_R'\gamma_2+t_R\gamma_3)\gamma_4+it_{LR}(\phi)\gamma_2\gamma_3+it_o\gamma_1\gamma_4\,,
\end{equation} 
where $t_{LR}(\phi)\equiv t_{LR}\cos{(\phi/2)}$ is the flux-dependent coupling between the $\gamma_2$ and $\gamma_3$ across the tunneling region, and $t_{L,R}$ ($t_{L,R}'$) are the coupling between the left (right) Majoranas. It is instructive to rewrite this Hamiltonian in terms of real fermions as follows:
\begin{align}
c_A&=\dfrac{1}{2}(\gamma_3+i\gamma_2);\,\,\,\,c_A^\dagger=\dfrac{1}{2}(\gamma_3-i\gamma_2)\,,\\
c_B&=\dfrac{1}{2}(\gamma_4+i\gamma_1);\,\,\,\,c_B^\dagger=\dfrac{1}{2}(\gamma_4-i\gamma_1)\,,
\end{align}
so that the Majorana operators can be written as:
\begin{align}
\gamma_1=-i(c_B-c_B^\dagger),\,\,\,\,\gamma_2=-i(c_A-c_A^\dagger),\,\,\,\,\gamma_3=c_A+c_A^\dagger,\,\,\,\,\gamma_4=c_B+c_B^\dagger\,.
\end{align}
In terms of the fermionic operators we obtain:

\begin{align}
H_M&=-t_{LR}(\phi)(2c_A^\dagger c_A-1)-t_{o}(2c_B^\dagger c_B-1)+i(t_R-t_L)(c_A^\dagger c_B-c_B^\dagger c_A)+i(t_R+t_L)(c_Ac_B-c_B^\dagger c_A^\dagger)\nonumber\\
&-(t_R'+t_L')(c_A^\dagger c_B+c_B^\dagger c_A)+(t_R'-t_L')(c_Ac_B+c_B^\dagger c_A^\dagger)\,.
\end{align}

Let us diagonalize this Hamiltonian to find the single particle Andreev levels. For that, we rewrite the many-body Hamiltonian in the basis $\vec{c}\equiv(c_A,c_B,c_A^\dagger,c_B^\dagger)^T$ as :
\begin{align}
H_M&=\dfrac{1}{2}\vec{c}^\dagger H_{A}\vec{c}\,,\\
H_A&=-(t_{LR}+t_o)\tau_z-(t_{LR}-t_o)\sigma_z\tau_z-(t_{R}-t_L)\sigma_y-(t_{R}+t_L)\sigma_y\tau_x-(t_L'+t_R')\sigma_x\tau_z-(t_L'-t_R')\sigma_y\tau_y\,,
\end{align}
where $\vec{\sigma}$ act in the $\{A,B\}$ basis, and $\vec{\tau}$ act in the particle-hole basis.

In order to diagonalize this Hamiltonian, we perform a series of unitary transformations $U(\theta_1,\theta_2,\theta_3,\theta_4)=\Pi_{i=1}^4U(\theta_i)$, that can lead to the following form:
\begin{align}
U^\dagger(\{\theta_i\})H_MU(\{\theta_i\})=-\left[\sqrt{(t_{LR}+t_0)^2+(t_{R}+t_L)^2+(t'_{R}-t'_L)^2}+\sqrt{(t_{LR}-t_0)^2+(t_{R}-t_L)^2+(t'_{R}+t'_L)^2}\sigma_z\right]\tau_z
\end{align}
while choosing
\begin{align}
U(\theta_1)&=e^{i\theta_1\sigma_y\tau_x/2}\,,\\
U(\theta_2)&=e^{i\theta_2\sigma_x\tau_z/2}\,,\\
U(\theta_3)&=e^{i\theta_3\sigma_y\tau_y/2}\,,\\
U(\theta_4)&=e^{i\theta_4\sigma_y/2}\,,
\end{align}
and 
\begin{align}
\theta_1&=\arctan{\frac{t_L'-t_R'}{t_{LR}+t_o}}\,,\\
\theta_2&=\arctan{\frac{t_R-t_L}{t_{LR}-t_o}}\,,\\
\theta_3&=\arctan{\frac{-t_R-t_L}{\sqrt{(t_{LR}+t_o)^2+(t'_R-t'_L)^2}}}\,,\\
\theta_4&=\arctan{\frac{-t'_R-t'_L}{\sqrt{(t_{LR}-t_o)^2+(t_R-t_L)^2}}}\,,
\end{align}
The single-particle spectrum of the system reads
\begin{align}
\epsilon_{1,\pm}(\phi)=\pm\left[\sqrt{(t_{LR}(\phi)+t_o)^2+(t_L+t_R)^2+(t'_R-t'_L)^2}+\sqrt{(t_{LR}(\phi)-t_o)^2+(t_L-t_R)^2+(t'_R+t'_L)^2}\right]\,,\\
\epsilon_{2,\pm}(\phi)=\pm\left[\sqrt{(t_{LR}(\phi)+t_o)^2+(t_L+t_R)^2+(t'_R-t'_L)^2}-\sqrt{(t_{LR}(\phi)-t_o)^2+(t_L-t_R)^2+(t'_R+t'_L)^2}\right]\,.
\end{align}
In the case of a symmetric wire, $t_{L}=t_R\equiv T$ and $t'_R=t'_L\equiv T'$. In this case, $\theta_1=\theta_2=0$, and we are left with only two rotations by angles:
\begin{align}
\theta_3&=-\arctan{\frac{2T}{t_{LR}+t_o}}\,,\\
\theta_4&=-\arctan{\frac{2T'}{t_{LR}-t_o}}\,,
\end{align}
and the energies
\begin{align}
\epsilon_{1,\pm}(\phi)=\pm\left[\sqrt{(t_{LR}(\phi)+t_o)^2+4T^2}+\sqrt{(t_{LR}(\phi)-t_o)^2+4T'^2}\right]\,,\\
\epsilon_{2,\pm}(\phi)=\pm\left[\sqrt{(t_{LR}(\phi)+t_o)^2+4T^2}-\sqrt{(t_{LR}(\phi)-t_o)^2+4T'^2}\right]\,.
\end{align}
From this, and assuming $t_o\approx0$, we can extract the magnitude of the anticrossing at $\phi=\pi$ to be $4T'$, while at $\phi=0$, the energy of highest state is $\approx2t_{LR}(0)$.  

Note that the single-particle current operator in the original basis is given by 
\begin{align}
\hat{j}\equiv\hat{I}_s=\frac{\partial t_{LR}}{\partial\Phi}(1+\sigma_z)\tau_z\,,
\end{align} 
since we assume $t_{LR}$ is the only quantity depending on $\Phi$ (the most important). We can define the current in the new transformed basis as $\hat{I}'_s=U^\dagger(\{\theta_i\})\hat{I}_sU(\{\theta_i\})$. The non-zero matrix elements of the current operator in the transformed basis read:

\begin{align}
j_{1,1}&=-j_{-1,-1}=\left(\cos{\theta_1}\cos{\theta_3}+\cos{\theta_2}\cos{\theta_4}\right)\frac{\partial t_{LR}(\phi)}{\partial\Phi}\,,\\
j_{2,2}&=-j_{-2,-2}=\left(\cos{\theta_1}\cos{\theta_3}-\cos{\theta_2}\cos{\theta_4}\right)\frac{\partial t_{LR}(\phi)}{\partial\Phi}\,,\\
j_{1,2}&=(j_{2,1})^*=-(j_{-1,-2})^*=-j_{-2,-1}=\left(i\sin{\theta_2}+\cos{\theta_2}\sin{\theta_4}\right)\frac{\partial t_{LR}(\phi)}{\partial\Phi}\,,\\
j_{1,-2}&=(j_{-2,1})^*=-(j_{-1,2})^*=-j_{2,-1}=\left(\sin{\theta_1}-i\cos{\theta_1}\sin{\theta_3}\right)\frac{\partial t_{LR}(\phi)}{\partial\Phi}\,,
\end{align}

while all the other matrix elements are zero. Note that for $t_L=t_R$ and $t'_L=t'_R$ (symmetric wire), we have  $\theta_{1}=\theta_2=0$, and the only matrix elements left are:

\begin{align}
j_{1,1}&=-j_{-1,-1}=\left(\cos{\theta_3}+\cos{\theta_4}\right)\frac{\partial t_{LR}(\phi)}{\partial\Phi}\,,\\
j_{2,2}&=-j_{-2,-2}=\left(\cos{\theta_3}-\cos{\theta_4}\right)\frac{\partial t_{LR}(\phi)}{\partial\Phi}\,,\\
j_{1,2}&=(j_{2,1})^*=-(j_{-1,-2})^*=-j_{-2,-1}=\sin{\theta_4}\frac{\partial t_{LR}(\phi)}{\partial\Phi}\,,\\
j_{1,-2}&=(j_{-2,1})^*=-(j_{-1,2})^*=-j_{2,-1}=-i \sin{\theta_3}\frac{\partial t_{LR}(\phi)}{\partial\Phi}\,,
\end{align}

Alternatively, we can also address the many-body spectrum of the Hamiltonian. The Hilbert space corresponding to the above Majorana Hamiltonian is spanned by four states, $\{|00\rangle,c_A^\dagger|00\rangle,c_B^\dagger|00\rangle,c_A^\dagger c_B^\dagger|00\rangle\}$, with $|00\rangle$ being the vacuum with no electrons. The general state can be written as $|n_An_B\rangle$, with $n_A=0,1$ and $n_B=0,1$; however, we need to pay attention at the ordering of the filling of the states.   This Hamiltonian conserves the total number of electrons (modulo $2$), and thus we can separate the full Hamiltonian in two diagonal blocks, for odd and even number of electrons, respectively. The odd subspace is spanned by the states $\{|01\rangle,|10\rangle\}$, while the even one by $\{|00\rangle,|11\rangle\}$. The resulting Hamiltonians in the odd and even subspaces, respectively, read:

\begin{align}
H_{M,o}&=\left(
\begin{array}{cc}
\langle 1_B 0_A| H_M |0_A 1_B\rangle & \langle 1_B 0_A| H_M |1_A 0_B\rangle\\
\langle 0_B 1_A| H_M |0_A 1_B\rangle & \langle 0_B 1_A| H_M |1_A 0_B\rangle
\end{array}
\right)=\left(
\begin{array}{cc}
t_{LR}(\phi)-t_o & -i(t_R-t_L)-(t'_R+t'_L)\\
i(t_R-t_L)-(t'_R+t'_L) & -t_{LR}(\phi)+t_o
\end{array}
\right)\,\\
H_{M,e}&=\left(
\begin{array}{cc}
\langle 0_B 0_A| H_M |0_A 0_B\rangle & \langle 0_B 0_A| H_M |1_A 1_B\rangle\\
\langle 1_B 1_A| H_M |0_A 0_B\rangle & \langle 1_B 1_A| H_M |1_A 1_B\rangle
\end{array}
\right)=\left(
\begin{array}{cc}
t_{LR}(\phi)+t_o & i(t_R+t_L)+(t'_R-t'_L)\\
-i(t_R+t_L)+(t'_R-t'_L) & -t_{LR}(\phi)-t_o
\end{array}
\right)\,.
\end{align} 

At this stage, it is also instructive to derive the current operator associated with this low energy Hamiltonian:
\begin{equation}
\hat{I}_s=-\frac{\partial H_{M}}{\partial \Phi}=\frac{\partial t_{LR}(\phi)}{\partial\Phi}(2c_A^\dagger c_A-1)\,.
\end{equation}

We can find the odd and even eigenvalues associated with the many-body odd and even Hamiltonians:
\begin{equation}
E_{\tau,\sigma}=\sigma\sqrt{(t_{LR}(\phi)+\tau t_o)^2+(t_L+\tau t_R)^2+(t'_L-\tau t'_R)^2}\,,
\end{equation}  
with $\sigma=\pm1$ and $\tau=\pm1\equiv e,o$. 
The single-particle energies can be found easily, by identifying the excitation spectrum.  They are given as follows:
\begin{align}
\epsilon_{1,\sigma}(\phi)=\sigma(E_{+,+}+E_{-,+})\,,\\
\epsilon_{2,\sigma}(\phi)=\sigma(E_{+,+}-E_{-,+})\,,
\end{align}
which correspond to the single particle energies found before.

\subsection{Low-energy susceptibility}

In the following we calculate the various components of the susceptibility for the cases without and with parity constraints. We start with the Josephson component.

\subsubsection{Josephson Susceptibility}

Without parity constraint, this reads
\begin{align}
\chi_J(\phi)&=-\sum_{n=1,2;\sigma}\frac{\partial}{\partial\Phi}\left[f(\epsilon_{n,\sigma})\frac{\partial\epsilon_{n,\sigma}}{\partial\Phi}\right]=-\sum_{n=1,2}\frac{\partial}{\partial\Phi}\left[\left(2f(\epsilon_{n})-1\right)\frac{\partial\epsilon_{n}}{\partial\Phi}\right]=\sum_{n=1,2}\frac{\partial}{\partial\Phi}\left[\tanh{(\beta\epsilon_{n}/2)}\frac{\partial\epsilon_{n}}{\partial\Phi}\right],
\end{align}
where $\epsilon_{n}\equiv\epsilon_{n,+}$ (positive energies).
This can be easily evaluated for both zero and finite temperatures, but let us give the expression in the former case. In this situation, $f(\epsilon_{n,+})=0$, and we are left with:
\begin{align}
\chi_J(\phi)&=\frac{\partial^2}{\partial\Phi^2}(\epsilon_1+\epsilon_2)\,.
\end{align}
We will not evaluate this any further, although it is very easy. We will focus instead on the parity constrained Josephson susceptibility. We get: 

\begin{align}
\chi_{\tau,J}(\phi)&=-\sum_{n=1,2}\frac{\partial}{\partial\Phi}\left[\left(2f_\tau(\epsilon_{n})-1\right)\frac{\partial\epsilon_{n}}{\partial\Phi}\right]\equiv \frac{\partial}{\partial\Phi}\left[\tanh{(\beta E_{+,\tau})}\frac{\partial E_{+,\tau}}{\partial\Phi}\right]\,.
\end{align}

\subsubsection{Diagonal Susceptibility}

The second term, which is the main term of interest here, is the diagonal one, which in the case of no parity constraint reads:
\begin{align}
\chi_D(\phi,\omega)&=\frac{-i\omega}{\gamma_D-i\omega}\sum_{n=1,2,\sigma}\frac{\partial f(\epsilon_{n,\sigma})}{\partial\Phi}\frac{\partial \epsilon_{n,\sigma}}{\partial\Phi}=\frac{-i\omega}{\gamma-i\omega}\sum_{n=1,2}\frac{\partial}{\partial\Phi}[2f(\epsilon_{n})-1]\frac{\partial\epsilon_{n}}{\partial\Phi}\nonumber\\
&=\frac{i\omega}{\gamma_D-i\omega}\left[\frac{\partial \tanh{(\beta\epsilon_{1}/2)}}{\partial\epsilon_{1}}\left(\frac{\partial\epsilon_{1}}{\partial\Phi}\right)^2+\frac{\partial \tanh{(\beta\epsilon_{2}/2)}}{\partial\epsilon_{2}}\left(\frac{\partial\epsilon_{2}}{\partial\Phi}\right)^2\right]\,,
\end{align}
where $\epsilon_{n}\equiv\epsilon_{n,+}$ (positive energies). This diagonal term is extremely sensitive to the presence of low-energy levels and on the temperature, and it can be easily calculated both analytically and numerically. 

Next we address the parity-constraint diagonal susceptibility.  This reads:

\begin{align}
\chi_{\tau,D}(\phi,\omega)&=-\frac{i\omega}{\gamma_D-i\omega}\sum_{n=1,2}\frac{\partial}{\partial\Phi}[2f_\tau(\epsilon_{n})-1]\frac{\partial\epsilon_{n}}{\partial\Phi}\,,
\end{align}
which when evaluated for each parity individually gives:
\begin{align}
\chi_{\tau,D}(\phi,\omega)&=\frac{2i\omega}{\gamma_D-i\omega}\frac{\partial \tanh{[\beta E_{+,\tau}]}}{\partial\Phi}\frac{\partial E_{+,\tau}}{\partial\Phi}=\frac{2i\omega}{\gamma_D-i\omega}\frac{\partial \tanh{[\beta E_{+,\tau}]}}{\partial E_{+,\tau}}\left(\frac{\partial E_{+,\tau}}{\partial\Phi}\right)^2\,,
\end{align}

Note that both in the Josephson and the diagonal component there the results for the two parities  depend on the $\epsilon_1\pm\epsilon_2\equiv 2E_{+,\pm}$, which are nothing but the many-body energies of the system associated with opposite parities. 

\subsubsection{Non-diagonal susceptibility}

Finally, the last term stands for the non-diagonal (or Kubo) contribution, and accounts for transitions between the Andreev levels. We recall the expression for this term (taking $\hbar=1$):
\begin{align}
\chi_{ND}(\phi,\omega)&=-\omega\sum_{n\neq m}\frac{f(\epsilon_n)-f(\epsilon_m)}{\epsilon_n-\epsilon_m}\frac{|\langle n|\hat{I}_s|m\rangle|^2}{\epsilon_n-\epsilon_m-\hbar\omega-i\hbar\gamma_{ND}}\nonumber\\
&=-2\omega\frac{f(\epsilon_{1})-f(\epsilon_{2})}{\epsilon_{1}-\epsilon_{2}}|\langle 1|\hat{I}_s|2\rangle|^2\frac{\omega+i\gamma_{ND}}{(\epsilon_{1}-\epsilon_{2})^2-(\omega+i\gamma_{ND})^2}\nonumber\\
&+2\omega\frac{1-f(\epsilon_{1})-f(\epsilon_2)}{\epsilon_{1}+\epsilon_2}|\langle 1|\hat{I}_s|-2\rangle|^2\frac{\omega+i\gamma_{ND}}{(\epsilon_{1}+\epsilon_2)^2-(\omega+i\gamma_{ND})^2}\nonumber\\
&=-2\omega\left(\frac{\partial t_{LR}(\phi)}{\partial\Phi}\right)^2\frac{f(\epsilon_{1})-f(\epsilon_{2})}{\epsilon_{1}-\epsilon_{2}}\frac{\omega+i\gamma_{ND}}{(\epsilon_{1}-\epsilon_{2})^2-(\omega+i\gamma_{ND})^2}\left(\sin^2{\theta_2}+\cos^2{\theta_2}\sin^2{\theta_4}\right)\nonumber\\
&+2\omega\left(\frac{\partial t_{LR}(\phi)}{\partial\Phi}\right)^2\frac{1-f(\epsilon_{1})-f(\epsilon_2)}{\epsilon_{1}+\epsilon_2}\frac{\omega+i\gamma_{ND}}{(\epsilon_{1}+\epsilon_2)^2-(\omega+i\gamma_{ND})^2}\left(\sin^2{\theta_1}+\cos^2{\theta_1}\sin^2{\theta_3}\right)\nonumber\\
&=-8\omega\left(\frac{\partial t_{LR}(\phi)}{\partial\Phi}\right)^2\Bigg[[(t_L-t_R)^2+(t'_L+t'_R)^2]\frac{f(\epsilon_{1})-f(\epsilon_{2})}{(\epsilon_{1}-\epsilon_{2})^3}\frac{\omega+i\hbar\gamma_{ND}}{(\epsilon_{1}-\epsilon_{2})^2-(\omega+i\gamma_{ND})^2}\nonumber\\
&-[(t_L+t_R)^2+(t'_L-t'_R)^2]\frac{1-f(\epsilon_{1})-f(\epsilon_2)}{(\epsilon_{1}+\epsilon_2)^3}\frac{\omega+i\gamma_{ND}}{(\epsilon_{1}+\epsilon_2)^2-(\omega+i\gamma_{ND})^2}\Bigg]\,,
\end{align}
where the first term  corresponds to scattering between states with the same quasiparticle number, while the last one to creating and annihilation of pairs of quasiparticles.  

Now we can also evaluate the parity-constrained non-diagonal susceptibility. This reads (from the above expression):
\begin{align}
\chi_{\tau,ND}(\phi,\omega)&=-8\omega\left(\frac{\partial t_{LR}(\phi)}{\partial\Phi}\right)^2\Bigg[[(t_L-t_R)^2+(t'_L+t'_R)^2]\frac{f_\tau(\epsilon_{1})-f_\tau(\epsilon_{2})}{(\epsilon_{1}-\epsilon_{2})^3}\frac{\omega+i\gamma_{ND}}{(\epsilon_{1}-\epsilon_{2})^2-(\omega+i\gamma_{ND})^2}\nonumber\\
&-[(t_L+t_R)^2+(t'_L-t'_R)^2]\frac{1-f_\tau(\epsilon_{1})-f_\tau(\epsilon_2)}{(\epsilon_{1}+\epsilon_2)^3}\frac{\hbar\omega+i\gamma_{ND}}{(\epsilon_{1}+\epsilon_2)^2-(\omega+i\gamma_{ND})^2}\Bigg],
\end{align}
or, after manipulating the expression:
\begin{align}
\chi_{\tau,ND}(\phi,\omega)&=\omega\left(\frac{\partial t_{LR}(\phi)}{\partial\Phi}\right)^2
[(t_L-\tau t_R)^2+(t'_L+\tau t'_R)^2]\frac{\tanh{[\beta E_{+,\tau}]}}{E_{+,\tau}^3}\frac{\omega+i\gamma_{ND}}{4E_{+,\tau}^2-(\omega+i\gamma_{ND})^2}\,.
\end{align}

Once again, this has a simple and very intuitive interpretation: in the presence of parity constraint, only the many-body levels with a given parity enter the expression for the susceptibility. 

\subsection{Comparison of the components}

\subsubsection{Unconstrained parity}

\begin{figure}[t] 
\centering
\includegraphics[width=0.7\linewidth]{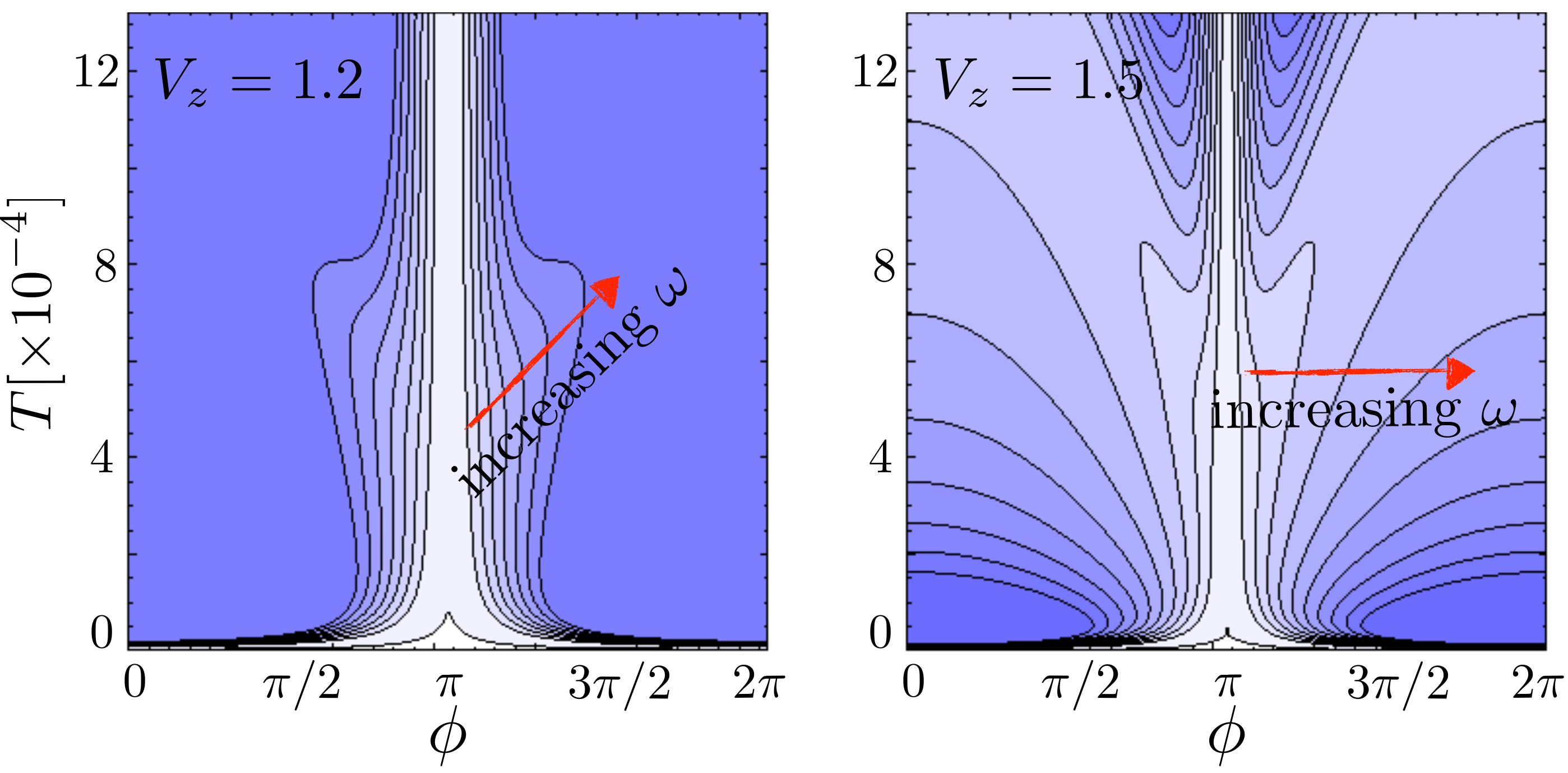}\
\caption{Left (Right): In blue, we  plot  the regions in which $\chi_{D}''(\phi,\omega)>\chi_{ND}''(\phi,\omega)$  as a function of $T$ and $\phi$ for different values of frequency $\omega=0,\omega_0, 2\omega_0\dots10\omega_0$, with $\omega_0=1.6\times10^{-4}$, and for $V_z=1.2\Delta_s^w$ ($V_z=1.5\Delta_s^w$) in the unconstrained parity case. The solid black lines correspond to the condition $\chi_{D}''(\phi,\omega)=\chi_{ND}''(\phi,\omega)$.  We expressed all energies in terms of the hopping $t$, with $\Delta_s^w=0.05$, $\alpha=0.08$,  and $\gamma_D=\gamma_{ND}=10^{-8}$.} 
\label{unconstrained} 
\end{figure}

We start with the unconstrained parity case. We focus only on the imaginary contribution, as  the part is dominated fully by the Josephson component. It is instructive to re-write the single-particle energy derivatives as follows:
\begin{align}
\frac{\partial\epsilon_{1,2}(\Phi)}{\partial\Phi}=\pm\frac{t_{LR}(\Phi)\epsilon_{1,2}}{E_{+,-}(\phi)E_{+,-}(\phi)}\frac{\partial t_{LR}(\Phi)}{\partial\Phi}\,.
\end{align} 
That allows us to write (the imaginary part) of $\chi_{D}$ as:
\begin{align}
\chi_D''(\phi,\omega)&=\frac{\omega\gamma_D}{\omega^2+\gamma_D^2}\left[\frac{\epsilon^2_1\partial \tanh{(\beta\epsilon_{1}/2)}}{\partial\epsilon_{1}}+\frac{\epsilon_2^2\partial \tanh{(\beta\epsilon_{2}/2)}}{\partial\epsilon_{2}}\right]\left(\frac{t_{LR}(\Phi)}{E_{+,+}E_{+,-}}\right)^2\left(\frac{\partial t_{LR}(\Phi)}{\partial\Phi}\right)^2\,.
\end{align}
On the other hand, we get for the imaginary part of the non-diagonal part:
\begin{align}
\chi''_{ND}(\phi,\omega)&=-\omega\gamma_{ND}\left(\frac{\partial t_{LR}(\phi)}{\partial\Phi}\right)^2\Bigg[[(t_L-t_R)^2+(t'_L+t'_R)^2]\frac{f(\epsilon_{1})-f(\epsilon_{2})}{E_{+,-}^3}\frac{4E_{+,-}^2+\omega^2+\gamma_{ND}^2}{(4E_{+,-}^2-\omega^2+\gamma_{ND}^2)^2+4\omega^2\gamma_{ND}^2}\nonumber\\
&-[(t_L+t_R)^2+(t'_L-t'_R)^2]\frac{1-f(\epsilon_{1})-f(\epsilon_2)}{E_{+,+}^3}\frac{4E_{+,+}^2+\omega^2+\gamma_{ND}^2}{(4E_{+,+}^2-\omega^2+\gamma_{ND}^2)^2+4\omega^2\gamma_{ND}^2}\Bigg]\,.
\end{align}
We get then for the ratio:
\begin{align}
\frac{\chi''_{ND}}{\chi''_{D}}&=\frac{\gamma_{ND}}{\gamma_D}(\omega^2+\gamma_D^2)\Bigg[F_-(T)\left(\frac{t_-}{t_{LR}}\right)^2\frac{(4E_{+,-}^2+\omega^2+\gamma_{ND}^2)}{(4E_{+,-}^2-\omega^2+\gamma_{ND}^2)^2+4\omega^2\gamma_{ND}^2}\nonumber\\
&+F_+(T)\left(\frac{t_+}{t_{LR}}\right)^2\frac{(4E_{+,+}^2+\omega^2+\gamma_{ND}^2)}{(4E_{+,+}^2-\omega^2+\gamma_{ND}^2)^2+4\omega^2\gamma_{ND}^2}\Bigg]\,,\nonumber\\
t_-&=\sqrt{(t_L-t_R)^2+(t'_L+t'_R)^2}\,,\\
t_+&=\sqrt{(t_L+t_R)^2+(t'_L-t'_R)^2}\,,\\
F_-(T)&=-\frac{\frac{f(\epsilon_{1})-f(\epsilon_{2})}{E_{+,-}}}{\frac{\epsilon^2_1}{E^2_{+,+}}\frac{\partial \tanh{(\beta\epsilon_{1}/2)}}{\partial\epsilon_{1}}+\frac{\epsilon^2_2}{E^2_{+,+}}\frac{\partial \tanh{(\beta\epsilon_{2}/2)}}{\partial\epsilon_{2}}}\,,\\
F_+(T)&=\frac{\frac{1-f(\epsilon_{1})-f(\epsilon_{2})}{E_{+,+}}}{\frac{\epsilon^2_1}{E^2_{+,-}}\frac{\partial \tanh{(\beta\epsilon_{1}/2)}}{\partial\epsilon_{1}}+\frac{\epsilon^2_2}{E^2_{+,-}}\frac{\partial \tanh{(\beta\epsilon_{2}/2)}}{\partial\epsilon_{2}}}\,,
\end{align}
where $T=1/\beta$. There are several limits that can be analyzed. However, we make the assumption that $\gamma_{D,ND}$ are the lowest energy scales over the entire parameter range. That is not necessary the case at ultra-low frequencies, but that is the typical experimental situation. We will assume frequencies $\omega\ll 2E_{+,\pm}$, $\omega\approx 2E_{+,\pm}$, as well as the case when $\omega\gg 2E_{+,\pm}$. Let us start with the former case. In such a situation, we get:
\begin{align}
\frac{\chi''_{ND}}{\chi''_{D}}&=\frac{\gamma_{ND}}{\gamma_D}\Bigg[F_-(T)\left(\frac{t_-}{t_{LR}}\right)^2\left(\frac{\omega}{2E_{+,-}}\right)^2+F_+(T)\left(\frac{t_+}{t_{LR}}\right)^2\left(\frac{\omega}{2E_{+,+}}\right)^2\Bigg]\,,
\end{align}
while for the resonant regime:
\begin{align}
\frac{\chi''_{ND}}{\chi''_{D}}&=\frac{\gamma_{ND}}{\gamma_D}F_{\mp}(T)\left(\frac{t_{\mp}}{t_{LR}}\right)^2\left(\frac{E_{+,\mp}}{\gamma_{ND}}\right)^2\,,
\end{align}
for $\omega=2E_{+,\mp}$ (and assuming that the off-resonant component is negligible). Finally, in the large frequency regime $\omega\gg 2E_{+,\mp}$ we obtain:
\begin{align}
\frac{\chi''_{ND}}{\chi''_{D}}&=\frac{\gamma_{ND}}{\gamma_D}\Bigg[F_-(T)\left(\frac{t_-}{t_{LR}}\right)^2+F_+(T)\left(\frac{t_+}{t_{LR}}\right)^2\Bigg]\,.
\end{align} 
In Fig.~\ref{unconstrained} we show the regions for which $\chi_D''(\phi,\omega)>\chi_{ND}''(\phi,\omega)$ as a function of $\phi$ and $T$, and for different values of $\omega$. We show the results for both $V_z=1.2\Delta_{s}^w$ and $V_z=1.5\Delta_{s}^w$. We see that there is a large region in the parameter space where indeed $\chi_D''(\phi,\omega)$ dominates over $\chi_{ND}''(\phi,\omega)$ and thus responsible for the dissipation, as claimed in the main text.

\begin{figure}[t] 
\centering
\includegraphics[width=0.7\linewidth]{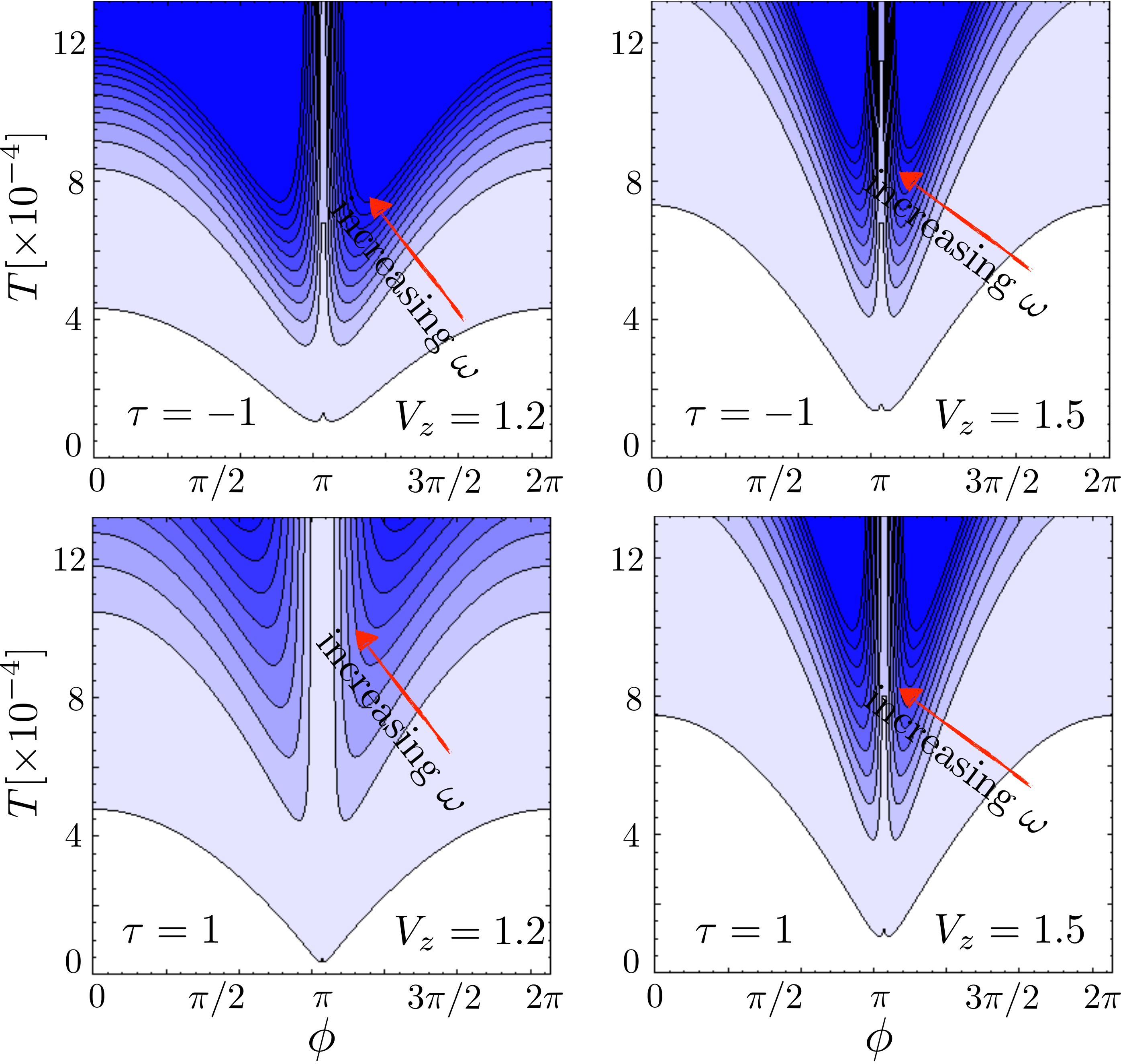}\
\caption{Left (Right): In blue we  plot the regions in which $\chi_{D}''(\phi,\omega)>\chi_{ND}''(\phi,\omega)$ as a function of $T$ and $\phi$ for different values of the frequencies $\omega$ (the same as in Fig.~\ref{unconstrained}), and  for $V_z=1.2\Delta_s^w$ ($V_z=1.5\Delta_s^w$) in the constrained parity case. Top (Bottom) plots are for parity $\tau=-1$ ($\tau=1$) case.  The solid black lines correspond to the condition $\chi_{D}''(\phi,\omega)=\chi_{ND}''(\phi,\omega)$.  We expressed all energies in terms of the hopping $t$, with $\Delta_s^w=0.05$, $\alpha=0.08$,  and $\gamma_D=\gamma_{ND}=10^{-8}$.} 
\label{constrained} 
\end{figure}

\subsubsection{Constrained parity}

Next we address the constrained parity situation.  In this case, the ratio becomes even simpler:
\begin{align}
\frac{\chi''_{\tau,ND}}{\chi''_{\tau,D}}&=\frac{\gamma_{ND}}{\gamma_D}\frac{\frac{\tanh{(\beta E_{+,\tau})}}{E_{+,\tau}}}{\frac{\partial \tanh{(\beta E_{+,\tau})}}{\partial E_{+,\tau}}}\left(\frac{t_\tau}{t_{LR}}\right)^2\frac{(\omega^2+\gamma_D^2)(4E_{+,\tau}^2+\omega^2+\gamma_{ND}^2)}{(4E_{+,\tau}^2-\omega^2+\gamma_{ND}^2)^2+4\omega^2\gamma_{ND}^2}\,,
\end{align}
which again can be analyzed in the cases $\omega\ll 2E_{+,\pm}$, $\omega\approx 2E_{+,\pm}$, and  $\omega\gg 2E_{+,\pm}$, respectively.  We get:
\begin{align}
\frac{\chi''_{\tau,ND}}{\chi''_{\tau,D}}&=\frac{\gamma_{ND}}{\gamma_D}\frac{\frac{\tanh{(\beta E_{+,\tau})}}{E_{+,\tau}}}{\frac{\partial \tanh{(\beta E_{+,\tau})}}{\partial E_{+,\tau}}}\left(\frac{t_\tau}{t_{LR}}\right)^2\left\{
\displaystyle{\begin{array}{cc}
\left(\frac{\omega}{2E_{+,\tau}}\right)^2 & {\rm for}\,\, \omega\ll2E_{+,\tau}\\
\left(\frac{E_{+,\tau}}{\gamma_{ND}}\right)^2 & {\rm for}\,\, \omega=2E_{+,\tau}\\
1 & {\rm for}\,\, \omega\gg 2E_{+,\tau}\,.
\end{array}}
\right.
\end{align}
In Fig.~\ref{constrained} we show the regions for which $\chi_{\tau,D}''(\phi,\omega)>\chi_{\tau,ND}''(\phi,\omega)$ as a function of $\phi$ and $T$, and for different values of $\omega$. We show the results for both $V_z=1.2\Delta_{s}^w$ and $V_z=1.5\Delta_{s}^w$. We see that there is a large region in the parameter space where indeed $\chi_D''(\phi,\omega)$ dominates over $\chi_{ND}''(\phi,\omega)$ and is thus responsible for the dissipation, as claimed in the main text.  However, the parameter range ($T,\omega,\phi$) over which the  diagonal term dominates is smaller than in the unconstrained case.


\begin{thebibliography}{39}
\bibitem [{\citenamefont {}\ \emph {}(NoteA)\citenamefont {},
  \citenamefont {}, \citenamefont {},\ and\
  \citenamefont {}}]{NoteA}%
  \BibitemOpen
\href@noop {} {\bibinfo  {journal} {The small ac component added to the magnetic flux threading
  the ring leads to a periodic driving $\delta \phi (t)$ of small amplitude
  around a fixed value of the phase $\phi $. This has to be contrasted with the
  linear time dependence that the phase acquires when biasing the junction with
  a dc voltage bias $V$, namely $\phi (t)=\protect \frac {2eV}{\hbar
  }t$}\
  }\BibitemShut {NoStop}%
\end{thebibliography}

\end{document}